\documentclass[%
reprint,
superscriptaddress,
showpacs,preprintnumbers,
amsmath,amssymb,
prb,
]{revtex4-1}
\usepackage{graphicx}% Include figure files
\usepackage{dcolumn}% Align table columns on decimal point
\usepackage{xfrac}
\usepackage{bm}% bold math
\usepackage{color}% text color
\usepackage[export]{adjustbox}
\usepackage{tabularx}
\newcommand{\RCL}{$\alpha$-RuCl$_3$}
\newcommand{\CIO}{Cu$_2$IrO$_3$}
\newcommand{\NIO}{Na$_2$IrO$_3$}
\newcommand{\LIO}{$\alpha$-Li$_2$IrO$_3$}

\newcommand{\ALIO}{Ag$_{3}$LiIr$_2$O$_6$}
\newcommand{\CLIO}{Cu$_{3}$LiIr$_2$O$_6$}
\newcommand{\ALSO}{Ag$_{3}$LiSn$_2$O$_6$}
\newcommand{\HLIO}{H$_{3}$LiIr$_2$O$_6$}

\newcommand{\C}{$^\circ$C}
\begin{document}
%\preprint{APS/123-QED}

\title{Effect of structural disorder on the Kitaev magnet \ALIO}

\author{Faranak~Bahrami}
\affiliation{Department of Physics, Boston College, Chestnut Hill, MA 02467, USA}
\author{Eric~M.~Kenney}
\affiliation{Department of Physics, Boston College, Chestnut Hill, MA 02467, USA}
\author{Chennan~Wang}
\affiliation{Laboratory for Muon Spin Spectroscopy, Paul Scherrer Institute, CH-5232 Villigen, Switzerland}
\author{Adam~Berlie}
\affiliation{ISIS Neutron and Muon Source, Science and Technology Facilities Council, Rutherford Appleton Laboratory, Didcot, OX11 0QX, United Kingdom}
\author{Oleg~I.~Lebedev}
\affiliation{Laboratoire CRISMAT, ENSICAEN-CNRS UMR6508, 14050 Caen, France}
\author{Michael~J.~Graf}
\affiliation{Department of Physics, Boston College, Chestnut Hill, MA 02467, USA}
\author{Fazel~Tafti}
\affiliation{Department of Physics, Boston College, Chestnut Hill, MA 02467, USA}
\email{fazel.tafti@bc.edu}

\date{\today}% It is always \today, today,
             %  but any date may be explicitly specified

\begin{abstract}
Searching for an ideal Kitaev spin liquid candidate with anyonic excitations and long-range entanglement has motivated the synthesis of a new family of intercalated Kitaev magnets such as \HLIO, \CIO, and \ALIO.
The absence of a susceptibility peak and a two-step release of the magnetic entropy in these materials has been proposed as evidence of proximity to the Kitaev spin liquid.
Here we present a comparative study of the magnetic susceptibility, heat capacity, and muon spin relaxation ($\mu$SR) between two samples of \ALIO\ in the clean and disordered limits.
In the disordered limit, the absence of a peak in either susceptibility or heat capacity and a weakly depolarizing $\mu$SR signal may suggest a proximate spin liquid ground state.
In the clean limit, however, we resolve a peak in both susceptibility and heat capacity data, and observe clear oscillations in $\mu$SR that confirm long-range antiferromagnetic ordering.
The $\mu$SR oscillations fit to a Bessel function, characteristic of an incommensurate order, as reported in the parent compound \LIO.
Our results clarify the role of structural disorder in the intercalated Kitaev magnets.  
\end{abstract}

\maketitle

%%%%%%%%%%%%%%%%%%%%%%%%%%%%%%%%%%%%%%%%%%%%%%%
\section{\label{sec:introduction}INTRODUCTION}
%%%%%%%%%%%%%%%%%%%%%%%%%%%%%%%%%%%%%%%%%%%%%%%
A long standing challenge in condensed matter physics has been to access a quantum spin liquid (QSL) ground state characterized by long-range entanglement and fractionalized anyonic excitations~\cite{broholm_quantum_2020,knolle_field_2019,savary_quantum_2016}.
One of the most promising theoretical models of QSL is the Kitaev model based on interacting spin-1/2 ions on a 2D honeycomb lattice with bond-dependent Ising axes~\cite{kitaev_anyons_2006}.
The prime candidates for the Kitaev model are \LIO, \NIO, and \RCL, but all three compounds order magnetically at low temperatures~\cite{jackeli_mott_2009,singh_relevance_2012,takagi_concept_2019,plumb_rucl_3_2014,nasu_fermionic_2016,wang_range_2020}.
Recently, a new class of intercalated Kitaev magnets have been synthesized via a topochemical exchange of the interlayer Li/Na atoms in \LIO\ and \NIO\ with H, Cu, or Ag atoms, and producing \HLIO, \CLIO, \CIO, and \ALIO~\cite{kitagawa_spinorbital-entangled_2018,roudebush_iridium_2016,abramchuk_cu2iro3_2017,bahrami_thermodynamic_2019}.  
It is claimed that this new family of Kitaev magnets, specifically \HLIO\ and \ALIO, are closer to the QSL phase based on the absence of magnetic ordering in thermodynamic measurements, a scaling behavior in the heat capacity, and a two-step release of the magnetic entropy~\cite{kitagawa_spinorbital-entangled_2018,kimchi_scaling_2018,knolle_bond-disordered_2019,bahrami_thermodynamic_2019}.
Both bond disorder and modified interlayer coordination are hypothesized as possible mechanisms for the proximity to the QSL ground-state~\cite{knolle_bond-disordered_2019,bahrami_thermodynamic_2019,choi_exotic_2019,abramchuk_cu2iro3_2017}.
Currently, there is no careful experimental work to examine these hypotheses and elucidate the role of structural disorder in the intercalated Kitaev magnets.

In this article, we present a careful study on the effect of structural disorder on one of the intercalated Kitaev magnets, \ALIO.
We show that the signatures of magnetic ordering may be hidden in a disordered sample, but they emerge unmistakably in a clean sample.  
Based on our experimental results, the onset of magnetic ordering in the clean limit is unaffected by the interlayer coordination, and the nature of disorder in \ALIO\ is inconsistent with a randomized bond picture~\cite{knolle_bond-disordered_2019}. 
Our experimental discussion is organized in four sections.
First, in a clean sample (S1), we reveal a broad peak in the magnetic susceptibility at the freezing temperature $T_F=14$~K followed by a sharp downturn at the N\'{e}el temperature $T_N=8$~K.
Such a peak and downturn are absent in a disordered sample (S2).
Second, we also reveal a peak in the heat capacity of S1 at $T_F$.
The peak turns into a mild change of slope in S2.
In the light of these findings, we will revisit the two-step entropy release that has been interpreted as evidence of spin fractionalization in \ALIO, \LIO, \NIO, and \RCL~\cite{bahrami_thermodynamic_2019,mehlawat_heat_2017,widmann_thermodynamic_2019}.
Third, by measuring the muon spin relaxation ($\mu$SR), we reveal short-range correlations at $T_F$ that turn into a long-range incommensurate order below $T_N$ in the clean sample S1.
The $\mu$SR data from sample S2 show slower depolarizations and less obvious oscillations compared to S1.
Fourth, we use transmission electron microscopy (TEM) to reveal extended regions of silver inclusion within the honeycomb layers of S2 that are absent in S1.
Complementary data and analyses are presented in four appendices at the end.

%%%%%%%%%%%%%%%%%%%%%%%%%%%%%%%%%%%%%%%%%%%%%%%
\section{\label{sec:experimental}EXPERIMENTAL METHODS}
%%%%%%%%%%%%%%%%%%%%%%%%%%%%%%%%%%%%%%%%%%%%%%%
We synthesized \ALIO\ via a topotactic cation-exchange reaction as reported in Ref.~\cite{bahrami_thermodynamic_2019}.
% %
% \begin{equation}
% \label{eq:topo}
% \mathrm{2Li_2IrO_3 + 3AgNO_3 \rightarrow Ag_3LiIr_2O_6 + 3LiNO_3}
% \end{equation}
% %
To improve the sample quality, however, we took two important additional measures.
First, we minimized the stacking faults in the precursor \LIO\ by performing a sequential solid-state synthesis at 900, 1000, and 1015~\C\ for 24, 32, and 48~h, respectively.
%In the first step, we pelletized a mixture of Li$_{2}$CO$_{3}$ (Alfa Aesar, 99.998\%) and iridium oxide (IrO$_{2}$, Alfa Aesar, 99\%) in the mole ratio 1.2:1 (with 20\% excess carbonate to account for its volatility at high temperatures).
%In subsequent steps, we did not add excess carbonate and sintered the pulverized instead of pelletized sample.
%The second improvement was to increase the duration of the exchange reaction.
%After several trials, we learned that Eq.~\ref{eq:topo} would complete at 350~\C\ after 24~h in a regular sample of \LIO, but a high-quality sample of \LIO\ would need at least one week to complete the reaction.
%
Second, we increased the duration of the topotactic reaction to several days to ensure a complete exchange of the high-quality \LIO\ precursor (see also Appendix~\ref{app:synthesis}). 
Sample S1 was made with the improved technique and sample S2 was made with the methods described in Ref.~\cite{bahrami_thermodynamic_2019}.

The electron diffraction (ED) and high angle annular dark field scanning TEM (HAADF--STEM) were performed using an aberration corrected JEM ARM200F microscope.
Powder X-ray diffraction (PXRD) was performed using a Bruker D8 ECO instrument equipped with a Cu-K$_\alpha$ source and a 1D LINXEYE-XE detector.
Magnetization and heat capacity were measured using Quantum Design MPMS3 and Dynacool PPMS, respectively.

The $\mu$SR experiments were carried at the Paul Scherrer Institute (PSI) using a $^3$He refrigerator with the Dolly Multi Purpose Surface-Muon Instrument (sample S1), and a gas flow cryostat with the General Purpose Surface-Muon (GPS) Instrument (both samples). 
Sample S1 was pressed into a pellet 13~mm in diameter and 1~mm thick, and sample S2 was 13~mm in diameter and 1.2~mm thick.
The pellets were wrapped in a 25~$\mu$m thin silver foil and mounted with varnish on copper holders. 
The same holder was used to mount S1 in both spectrometers.
Initial measurements were made on sample S2 using a dilution refrigerator and gas flow cryostat on the EMU spectrometer at the ISIS Muon Source at the Rutherford Appleton Laboratory.

%%%%%%%%%%%%%%%%%%%%%%%%%%%%%%%%%%%%%%%%%%%%%%%
\section{\label{sec:results}RESULTS AND DISCUSSION}
%%%%%%%%%%%%%%%%%%%%%%%%%%%%%%%%%%%%%%%%%%%%%%%
\subsection{\label{subsec:magnetization}Magnetic Susceptibility}
%%%%%%%%%%%%%%%%%%%%%%%%%%%%%%%%%%%%%%%%%%%%%%%
\begin{figure}
\includegraphics[width=0.46\textwidth]{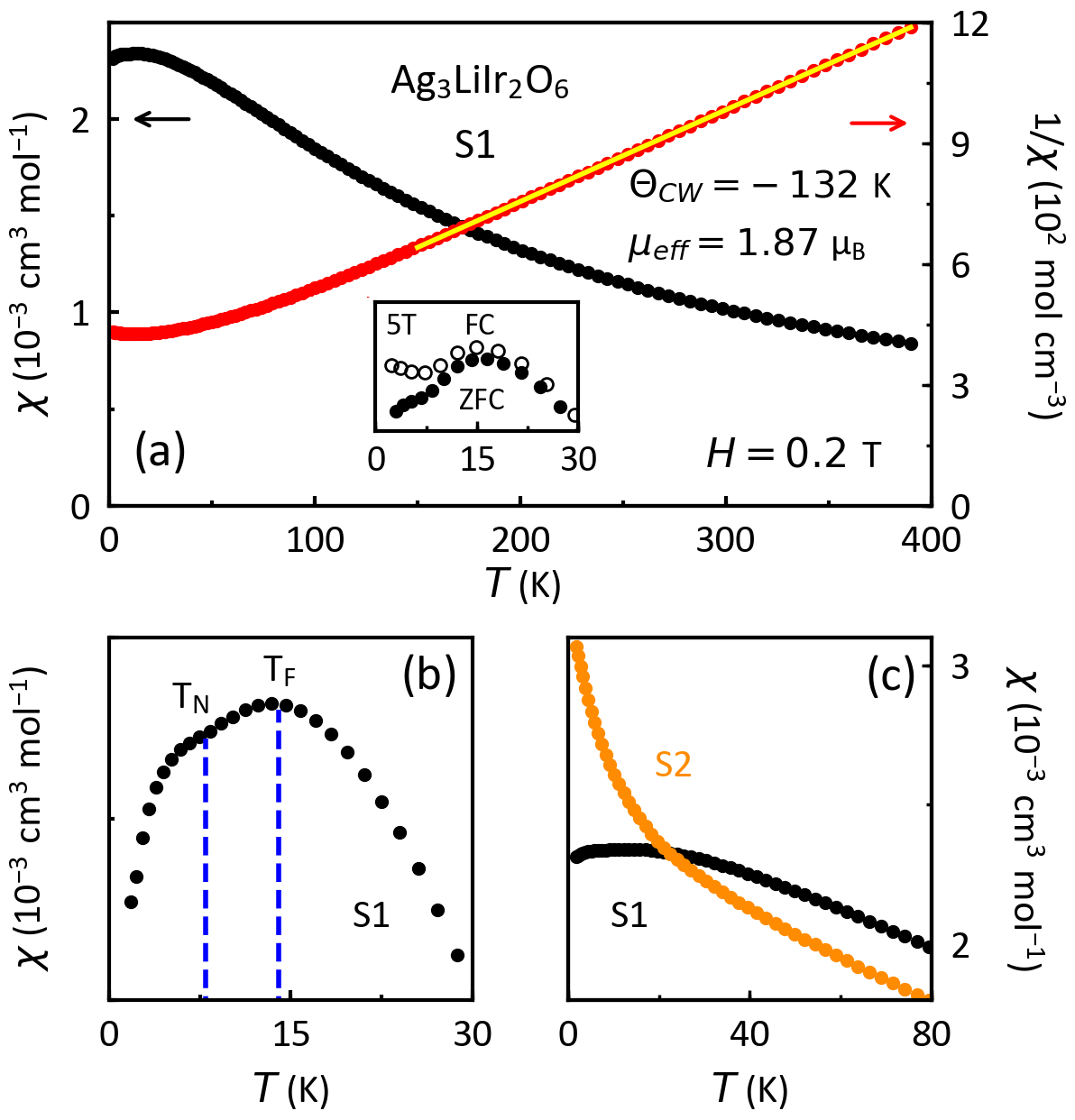}
\caption{\label{fig:CW}
(a) DC magnetic susceptibility per mole Ir (black data) and inverse susceptibility (red data) plotted as a function of temperature in the high-quality sample S1.
The yellow line is a Curie-Weiss fit at $T>150$~K.
The full and open circles in the inset represent the zero-field-cooled (ZFC) and field-cooled (FC) curves at $H=5$~T.
(b) Magnified view of the ZFC susceptibility from sample S1 showing a broad peak at $T_F=14$~K and a sharper downturn at $T_N=8$~K.
(c) $\chi(T)$ curves are compared between the clean sample S1 (black points) and disordered sample S2 (orange data from Ref.~\cite{bahrami_thermodynamic_2019}).
}
\end{figure}
%%%%%%%%%%%%%%%%%%%%%%%%%%%%%%%%%%%%%%%%%%%%%%%
The first evidence of magnetic ordering in a high-quality \ALIO\ sample (S1) is a peak in the DC susceptibility ($\chi$) as seen in Fig.~\ref{fig:CW}(a) and magnified in Fig.~\ref{fig:CW}(b).
The peak is broad and it splits between the zero-field-cooled (ZFC) and field-cooled (FC) conditions.
The inset of Fig.~\ref{fig:CW}(a) shows a persistent ZFC/FC splitting at 5~T which is consistent with a static spin freezing~\cite{raju_magnetic-susceptibility_1992,mydosh_spin_2015}, hence the label $T_F$ at 14~K in Fig.~\ref{fig:CW}(b). 
We show several ZFC/FC curves at different fields in Appendix~\ref{app:zfc}.
There is a second temperature scale $T_N$ at 8~K in Fig.~\ref{fig:CW}(b), below which the susceptibility turns down visibly (and the $\mu$SR data reveal clear oscillations in Section~\ref{subsec:muSR}).
Thus, we assign $T_N=8$~K to a long-range antiferromagnetic (AFM) order, and $T_F=14$~K to the onset of short-range static magnetism.

We compare the magnetic susceptibility of the clean sample (S1) and disordered sample (S2) in Fig.~\ref{fig:CW}(c).
A susceptibility peak is present in the former, but absent in the latter.
The absence of such a peak in a sample with the same quality as S2 has been misinterpreted as evidence of proximity to a Kitaev spin liquid~\cite{bahrami_thermodynamic_2019}.
After tremendous efforts to remove disorder and improve the quality of \ALIO, we were able to resolve the AFM peak in the high-quality sample S1.
Based on our results, it would be insightful to revisit recent claims of a quantum spin-liquid phase in another Kitaev material \HLIO\ which suffers from a higher disorder level than \ALIO~\cite{kitagawa_spinorbital-entangled_2018,bette_solution_2017}. 
A large low-temperature tail in $\chi(T)$ has been observed in \HLIO\ similar to the behavior of sample S2 in Fig.~\ref{fig:CW}(c).
The question is whether a peak is hidden under that low-temperature tail.
In a similar vein, recent claims of a disordered QSL phase in \CIO\ based on the absence of a peak in $\chi(T)$ may be questionable~\cite{choi_exotic_2019}.
In fact, a small peak at 2~K has been reported in higher-quality samples of that material and diagnosed as a signature of partial static magnetism~\cite{kenney_coexistence_2019}.

To understand the magnetic interactions in \ALIO, we performed a Curie-Weiss (CW) analysis on the inverse susceptibility ($1/\chi$) in Fig.~\ref{fig:CW}(a).
The yellow line represents the CW fit that yields a CW temperature $\Theta_{\textrm{CW}}=-132(1)$~K and a magnetic moment $\mu_{\textrm{eff}}=1.87(2)$~$\mu_{B}$.
The negative sign of $\Theta_{\textrm{CW}}$ indicates AFM interactions and its large magnitude, compared to $T_F$, implies magnetic frustration~\cite{ramirez_strongly_1994}.
We extract an effective magnetic moment of $\mu_{\textrm{eff}}=1.87$~$\mu_{B}$ from the CW fit which is comparable to the reported values in other Kitaev magnets~\cite{abramchuk_cu2iro3_2017,mehlawat_heat_2017} and close to the expected moment for a J$_{\textrm{eff}}=1/2$ system (1.74~$\mu_B$).
%
% In a recent work on another honeycomb material CdIrO$_3$, a deviation of the magnetic moment from the ideal J$_{\textrm{eff}}=1/2$ value has been attributed to the trigonal distortion of the octahedral coordination around each Ir$^{4+}$ ion~\cite{haraguchi_strong_2020}.
% %
% A trigonal distortion mixes the J$_{\textrm{eff}}=1/2$ and $3/2$ states and yields an effective moment slightly larger than a purely J$_{\textrm{eff}}=1/2$ state.
% %
% Note that this effect is relatively weak in \ALIO, but for completeness, we analyze the trigonal distortion in Appendix~\ref{app:trigonal}.
% %
The values of $\mu_{\textrm{eff}}$ and $\Theta_{\textrm{CW}}$ are comparable between S1 ($1.87~\mu_{B}$, $-132$~K) and S2 ($1.79~\mu_{B}$, $-142$~K)~\cite{bahrami_thermodynamic_2019}.

%%%%%%%%%%%%%%%%%%%%%%%%%%%%%%%%%%%%%%%%%%%%%%%
\subsection{\label{subsec:heat}Heat Capacity}
%%%%%%%%%%%%%%%%%%%%%%%%%%%%%%%%%%%%%%%%%%%%%%%
\begin{figure}
\includegraphics[width=0.46\textwidth]{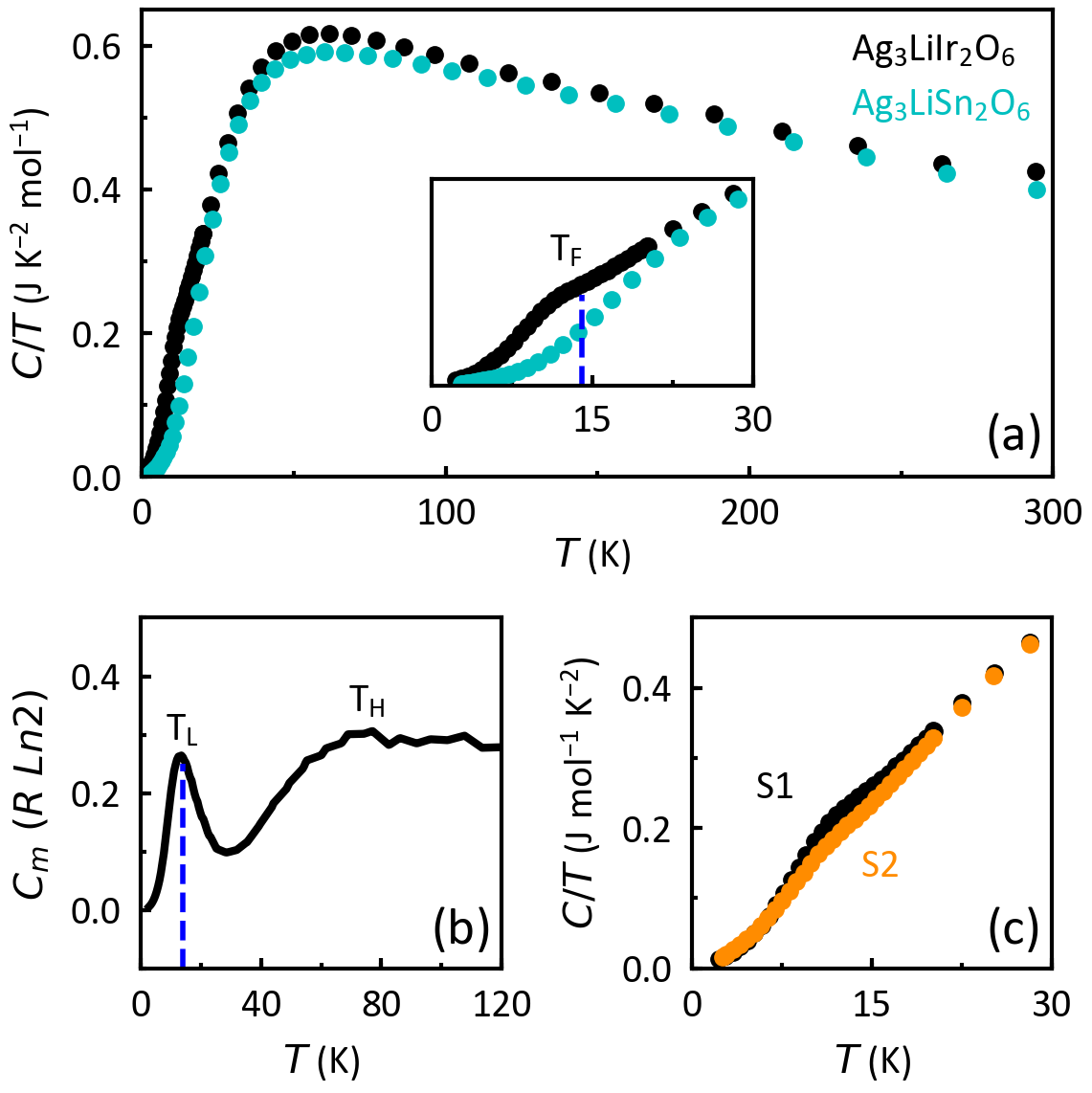}
\caption{\label{fig:HC}
(a) Heat capacity divided by temperature ($C/T$) per mole Ir or Sn plotted as a function of temperature in \ALIO~(black data) and its nonmagnetic lattice model \ALSO~(turquoise data from ref.~\cite{bahrami_thermodynamic_2019}).
(b) Magnetic heat capacity ($C_m$) in units of $R\ln(2)$ as a function of temperature below 120~K in S1, where $T_L=T_F=14$~K and $T_H=75$~K.
(c) Comparison between $C/T$ as a function of temperature below 30 K in the clean sample S1~(black) and disordered sample S2 (orange). 
The orange curve (from Ref.~\cite{bahrami_thermodynamic_2019}) is shifted by $-0.014$~K for clarity.
}
\end{figure}
%%%%%%%%%%%%%%%%%%%%%%%%%%%%%%%%%%%%%%%%%%%%%%%
We measured the heat capacity ($C$) of sample S1 to confirm the bulk AFM order in \ALIO.
Figure~\ref{fig:HC}(a) shows a broad peak in $C/T$ at $T_F=14$~K, reminiscent of the broad peak at $T_F$ in $\chi(T)$.
On the same figure, we also present the heat capacity of an isostructural compound \ALSO\ which serves as a non-magnetic lattice model for \ALIO.
The two data sets closely track each other as a function of temperature, except near 75~K and 14~K, where an additional magnetic contribution enhances the heat capacity of \ALIO.
The magnetic heat capacity ($C_m$) can be isolated by subtracting the \ALSO\ data from \ALIO.
Figure~\ref{fig:HC}(b) shows $C_m$ in units of $R\ln(2)$ as a function of temperature where two broad peaks are resolved at $T_H=75$~K and $T_L=14$~K.
Such a behavior has been interpreted as evidence of a fractionalization of spins into Majorana fermions at $T_H$ followed by a long-range entanglement at $T_L$ in \ALIO, \LIO, \NIO, and \RCL~\cite{mehlawat_heat_2017,widmann_thermodynamic_2019,bahrami_thermodynamic_2019}, based on a quantum Monte Carlo simulation of the Kitaev Hamiltonian~\cite{nasu_thermal_2015}. 
We caution against this interpretation and point out that the peak at $T_L$=$T_F$ in \ALIO\ is due to static magnetism instead of quantum entanglement. 
%More specifically, $T_F=T_L$ marks the onset of short-correlations which become long-range below $T_N$.
%The behavior of sample S2 with a peak in $C_m$ but without a peak in $\chi(T)$ is especially misleading as it indicates an entropy release without magnetic ordering.

We compare the $C/T$ curves between samples S1 (clean) and S2 (disordered) in Fig.~\ref{fig:HC}(c).
Whereas S2 shows a slight change of slope at $T_F=14$~K, S1 reveals a peak.
Notice that without having the clean sample S1, the heat capacity of S2 could have been misinterpreted as the absence of magnetic ordering.
This shows the importance of improving sample quality, since without having access to S1, we could not have associated the peak at $T_L$ with the entropy release from a long-range AFM oder instead of entanglement.
Similarly, the low temperature peaks in the heat capacity of \LIO, \NIO, and \RCL\ are due to AFM ordering~\cite{mehlawat_heat_2017,widmann_thermodynamic_2019,bahrami_thermodynamic_2019}.

The above discussion does not discredit the iridate materials as candidates of a Kitaev spin liquid.
Note that the peak at $T_H$ may indeed signal the onset of a fractionalization process, but the Majorana liquid develops an instability toward a gapped AFM state instead of melting into an entangled spin liquid ground state. 
In \RCL, this instability is removed by applying a 7~T magnetic field parallel to the honeycomb planes~\cite{banerjee_excitations_2018}.
A similar effect may be observed in \ALIO\ once single crystals are available.

%%%%%%%%%%%%%%%%%%%%%%%%%%%%%%%%%%%%%%%%%%%%%%%
\subsection{\label{subsec:muSR}Muon Spin Relaxation ($\mu$SR)}
%%%%%%%%%%%%%%%%%%%%%%%%%%%%%%%%%%%%%%%%%%%%%%%
\begin{figure}
\includegraphics[width=0.46\textwidth]{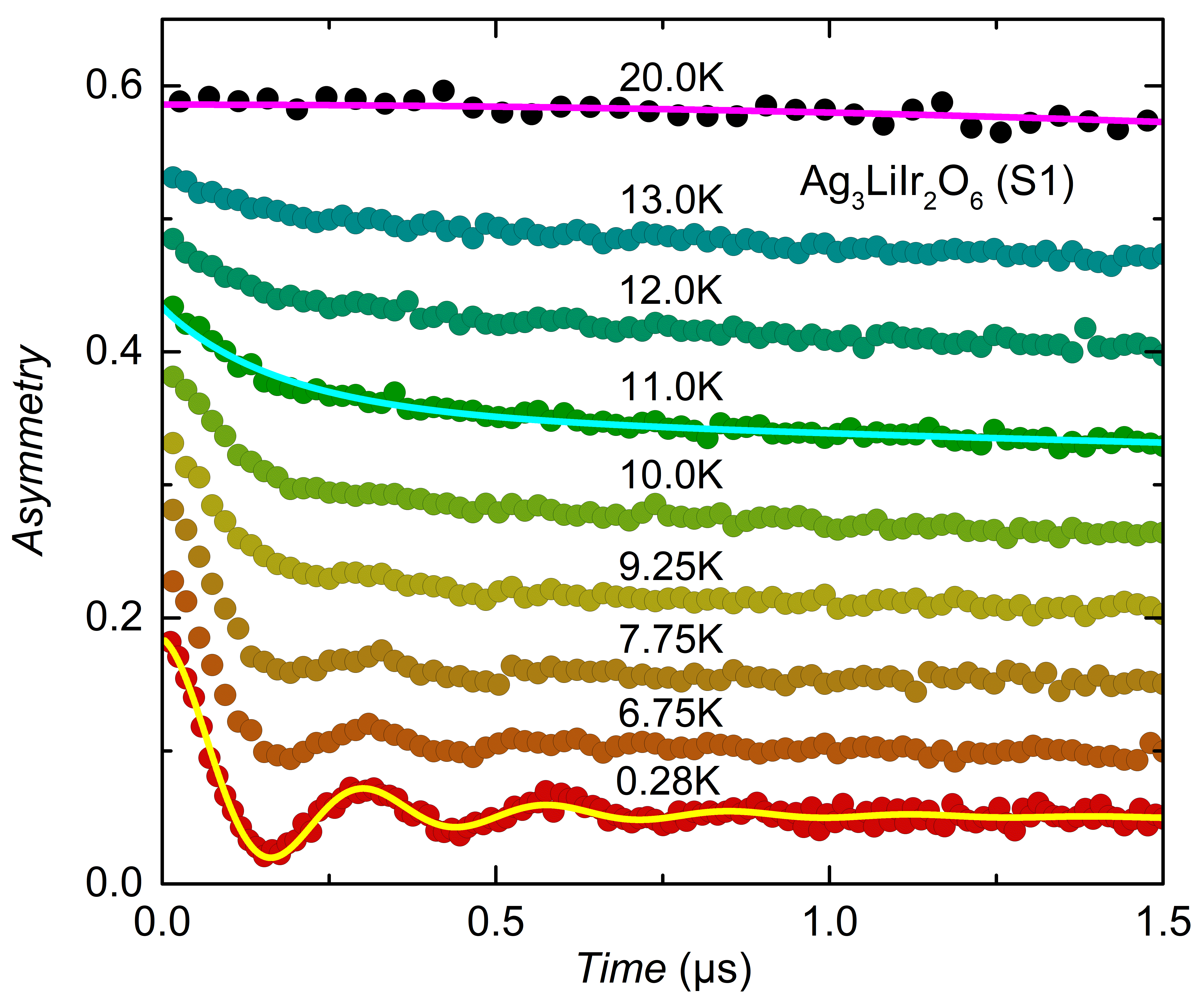}
\caption{\label{fig:TSWEEP}
Asymmetry plotted as a function of time at short timescales. 
The curves have been offset by equal increments from the base-temperature curve (0.28~K) for clarity.
The magenta, cyan, and yellow solid lines are fits to Eq.~\ref{eq:KT}, Eq.~\ref{eq:twochannel}, and Eq.~\ref{eq:bessel}, respectively.
}
\end{figure}
%%%%%%%%%%%%%%%%%%%%%%%%%%%%%%%%%%%%%%%%%%%%%%%
In positive muon spin relaxation ($\mu^+$SR), spin-polarized positive muons are injected into a sample and in less than 1~ps come to rest at a preferred crystallographic interstitial site (or sites). 
The muon spin polarization then evolves with time in the local magnetic field, yielding information about the magnitude and orientation of the local field relative to the initial spin direction.
After tens of millions of decay events, a time histogram can be used to extract the asymmetry, which is proportional to the time dependence of the projection of the muon spin along the detector direction~\cite{Yaouanc_muon_2011}.
The asymmetry contains information about the local field’s temporal and spatial variation.

We plot the asymmetry as a function of time in sample S1 in Fig.~\ref{fig:TSWEEP}, at nine representative temperatures from 0.28 to 20~K at zero field.
For temperatures greater than or equal to 20~K, the Ir$^{4+}$ moments are fluctuating too rapidly, and they have no effect on the muon.
Therefore, the depolarization is dominated by randomly oriented quasistatic nuclear moments. 
The temperature dependent asymmetry is well described by a Gaussian Kubo-Toyabe function
\begin{equation}
\label{eq:KT}
G_{\textrm{KT}}(t)=A_0\left[\frac{1}{3} + \frac{2}{3}\left( 1-\sigma^2 t^2 \right) \exp\left(-\frac{1}{2}\sigma^2 t^2\right) \right]
\end{equation}
where $A_0=0.174$ is the initial asymmetry for GPS in spin-rotated mode, and the parameter $\sigma=0.150$~MHz is proportional to the second moment of the field distribution experienced by the muon ensemble. 
The magenta line on Fig.~\ref{fig:TSWEEP} is a fit to Eq.~\ref{eq:KT} at 20~K.
We found a constant value for $\sigma$ between 200 and 20~K, indicating that the muon is not diffusing in this temperature range. 
The data below 20~K can be explained in three regions of interest.

\emph{Region 1.} For 20~K$>T>T_F$, depolarization is dominated by the nuclear moments.
The electronic moments are slowing down and begin contributing to muon depolarization.

\emph{Region 2.} For the range $T_F>T>T_N$, depolarization is dominated by the electronic moments. 
The spin freezing is manifest in the onset of a fast relaxation component in addition to a slow exponential depolarization due to fluctuations. 
To characterize the crossover in this temperature range, we use a phenomenological depolarization function
\begin{equation}
\label{eq:twochannel}
G(t)=A_0\left[\alpha_F \exp\left(-\left(\lambda_F t\right)^\beta\right) + \left(1-\alpha_F\right) \exp\left(-\lambda_St\right) \right]
\end{equation}
where $A_0=0.185$ is the initial asymmetry in the Dolly spectrometer in spin-rotated mode. 
The first term in the brackets is related to the fast decay with rate $\lambda_F$ best described by a stretched exponential with exponent $\beta$, and attributed to spin freezing. 
The second term is a slow exponential decay at rate $\lambda_S$ attributed to a fluctuating contribution. 
The fit parameters $\lambda_F$, $\beta$, and $\lambda_S$ in sample S1 vary from 10.1(6)~$\mu\textrm{s}^{-1}$, 0.85(6), and 0.211(2)~$\mu\textrm{s}^{-1}$ at 13~K, respectively, to 11.0(1)~$\mu\textrm{s}^{-1}$, 1.75(5), and 0.285(8)~$\mu\textrm{s}^{-1}$ at 8~K.

\begin{figure}
\includegraphics[width=0.46\textwidth]{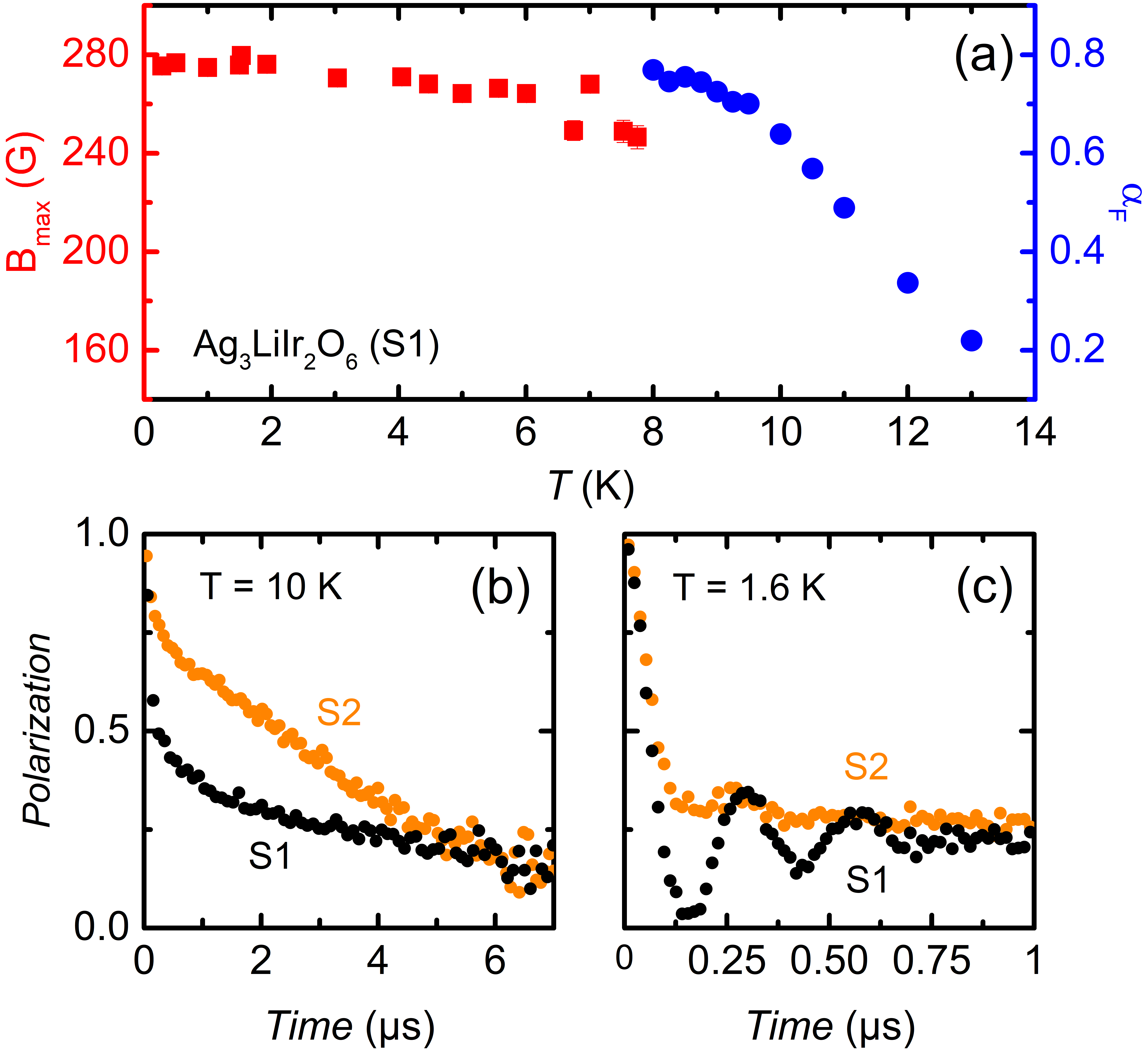}
\caption{\label{fig:OP}
(a) The blue circles represent $\alpha_F$ values from fits to Eq.~\ref{eq:twochannel}, and the red squares represent $B_\textrm{max}$ values from fits to Eq.~\ref{eq:bessel} in the clean sample S1.
Static magnetism starts at $T_F=14$~K and $\mu$SR oscillations start at $T_N=8$~K.
(b) Muon polarization ($P=A/A_0$ where $A_0$ is the initial asymmetry) as a function of time in S1 and S2 at 10~K ($T_F>T>T_N$).
(c) Polarization curves below 1~$\mu$s in S1 and S2 at 1.6~K ($T<T_N$). 
The oscillations are barely discernible in the disordered sample S2, although the initial depolarization is comparable between S1 and S2.
}
\end{figure}

The cyan line on Fig.~\ref{fig:TSWEEP} is a representative fit to Eq.~\ref{eq:twochannel} at 11~K.
From such fits, we extract the fraction of fast decay $\alpha_F$, which we take as a metric for the onset of static magnetism.
The temperature dependence of $\alpha_F$ is plotted in Fig.~\ref{fig:OP}(a), and it vanishes near $T_F=14$~K.

We compare the polarization (normalized asymmetry) at 10~K between samples S1 and S2 in Fig.~\ref{fig:OP}(b).
At this temperature ($T_F>T>T_N$), neither S1 nor S2 shows oscillations; however, the fast decay below 1~$\mu$s is visibly faster in S1.
Note that the long-time tail of polarization converges to the same value in both samples, indicating weak dynamics. 
We conclude that both samples undergo spin freezing below $T_F$, but the short-range correlations are stronger in the clean sample S1 as evidenced by a larger $\lambda_F$ than in the disordered sample S2.

%%%%%%%%%%%%%%%%%%%%%%%%%%%%%%%%%%%%%%%%%%%%%%%
%%%%%%%%%%%%%%%%%%%%%%%%%%%%%%%%%%%%%%%%%%%%%%%

\emph{Region 3.} At $T<T_N$, clear oscillations appear in the depolarization curves of S1 (Fig.~\ref{fig:TSWEEP}), indicating a long-range magnetic order.
The depolarization curves are well described by the function 
\begin{eqnarray}
\label{eq:bessel}
G_{\textrm{LRO}}(t)&=A_0[\alpha_{\textrm{LRO}} \exp\left(-\Lambda t\right)J_0\left(\gamma_\mu B_{\textrm{max}}t+\phi\right) \nonumber \\
 &+\left(1-\alpha_{\textrm{LRO}}\right)\exp\left(-\lambda t\right)]
\end{eqnarray}
Again, the initial asymmetry is $A_0=0.185$ in the Dolly spectrometer. 
Here $J_0$ is the zeroth-order Bessel function and the muon gyromagnetic ratio is $\gamma_\mu=2\pi(135.5~\textrm{MHz/T})$. 
The yellow line on Fig.~\ref{fig:TSWEEP} is a fit to the Bessel function at 0.28~K.
A Bessel oscillatory behavior is typically associated with incommensurate magnetic ordering~\cite{Yaouanc_muon_2011}, where the muon experiences ordered fields ranging from 0 to $B_{\textrm{max}}$. 
We extract the $B_{\textrm{max}}$ value from such a fit at each temperature below $T_N$, and plot it in Fig.~\ref{fig:OP}(a) as red squares.
Such an analysis would be impossible for the disordered sample S2 as can be seen from the comparison in Fig.~\ref{fig:OP}(c).
The oscillations are barely visible in S2, thus a fit to Eq.~\ref{eq:bessel} would not work. 
Two additional observations in Fig.~\ref{fig:OP}(c) are worth noting.
First, at extremely short timescale (less than 0.1~$\mu$s), the fast depolarization is identical in both samples.
Second, the long-time depolarization tail ($t>0.8~\mu$s) converges between the two samples.
From these observations, we conclude that a similar incommensurate order occurs in both samples below $T_N$, but with a longer correlation length in sample S1 than in S2, due to less disorder.

At the base temperature $T=0.28$~K, the fit to Eq.~\ref{eq:bessel} yields $\alpha_{\mathrm{LRO}}=0.741(2)$, $B_{\mathrm{max}}=269(1)$~G, $\phi =-0.9(6)^\circ$, $\Lambda =2.8(1)~\mathrm{\mu s}^{-1}$, and $\lambda=0.052(4)~\mathrm{\mu s}^{-1}$. 
The value for $\alpha_{\mathrm{LRO}}$ is close to the value 2/3 expected from a polycrystalline sample exhibiting long-range magnetic order. 
The value for $B_{\mathrm{max}}$ is confirmed from a longitudinal field (LF) experiment in Appendix~\ref{app:muH}.
The damping rate $\lambda$ is associated with those muons whose initial polarization lies along the local magnetic field and are depolarized by transverse magnetic fluctuations.
The rate $\Lambda$ contains contributions from both static magnetic disorder and magnetic fluctuations.
Since $\Lambda \gg \lambda$, disorder is the dominant contribution.

%%%%%%%%%%%%%%%%%%%%%%%%%%%%%%%%%%%%%%%%%%%%%%%
\subsection{\label{subsec:tem}Transmission Electron Microscopy (TEM)}
%%%%%%%%%%%%%%%%%%%%%%%%%%%%%%%%%%%%%%%%%%%%%%%
\begin{figure}
\includegraphics[width=0.46\textwidth]{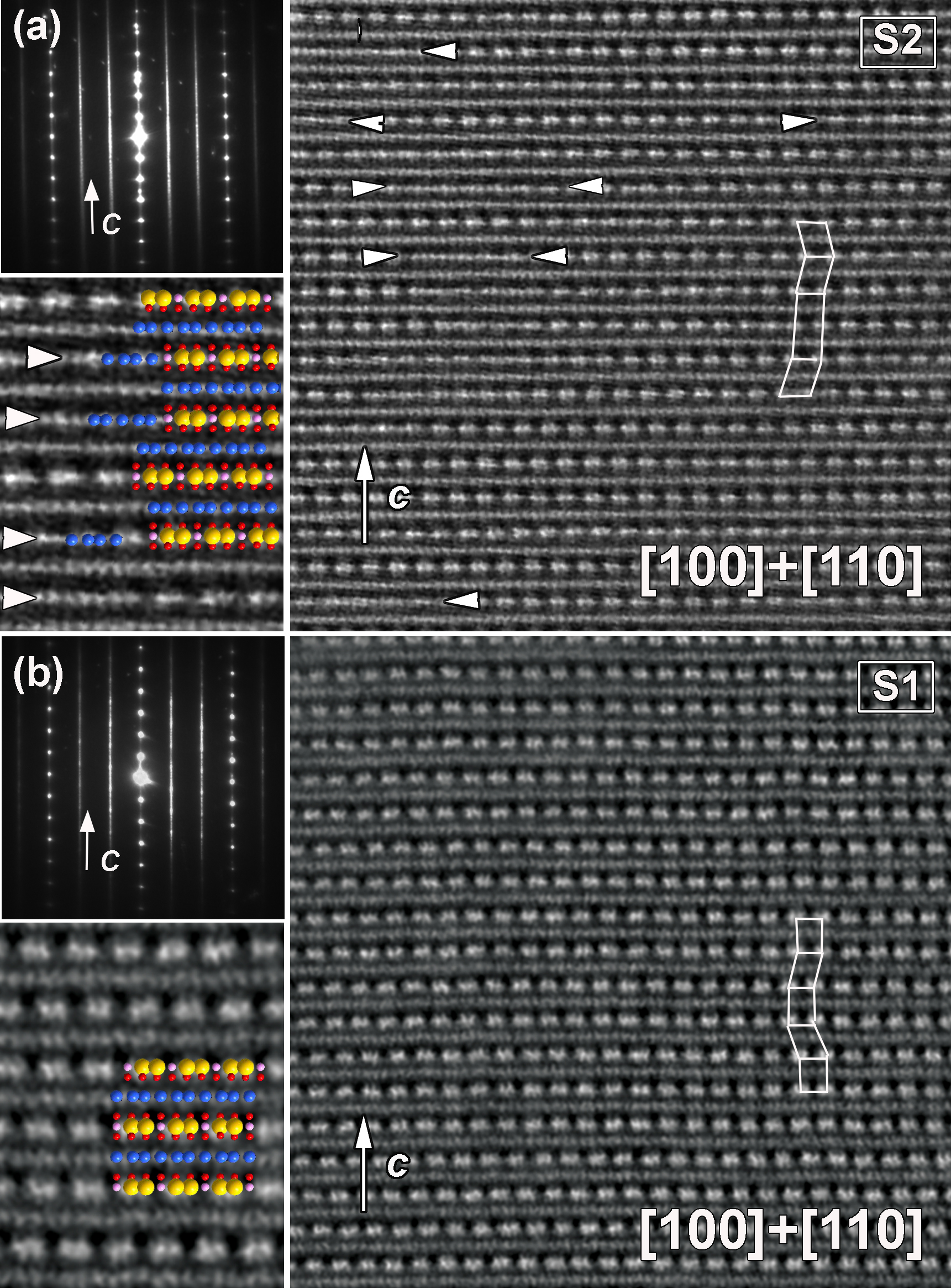}
\caption{\label{fig:TEM}
(a) Electron diffraction (top inset) and HAADF-STEM image from the disordered sample S2.
A structural model is overlaid on the magnified image in the bottom inset with blue, yellow, pink, and red spheres for the Ag, Ir, Li, and O atoms, respectively.
The arrows indicate where Ag atoms replace Ir atoms within the honeycomb layers.
(b) Similar images from the clean sample S1 where Ag inclusion is absent.
}
\end{figure}
%%%%%%%%%%%%%%%%%%%%%%%%%%%%%%%%%%%%%%%%%%%%%%%
So far, we have presented the magnetic behavior of \ALIO\ in the clean (S1) and disordered (S2) limits using both bulk and local probes.
Here we characterize the structural disorder in the material using high-resolution HAADF-STEM images from both samples S1 and S2 in Fig.~\ref{fig:TEM}.
%The images reveal extended regions of silver inclusion within the honeycomb layers in S2, unlike S1 which is devoid of such inclusions.
%
The characteristic feature of each honeycomb layer in Fig.~\ref{fig:TEM}(a,b) is a repeating pattern of a pair of Ir atoms (large bright spots) separated by a Li atom (not visible).
This pattern is interrupted in sample S2 by rows of unwanted Ag atoms (smaller bright spots) as indicated by the arrows in Fig.~\ref{fig:TEM}(a). 
Note that silver inclusions take the form of extended defects (rows of Ag atoms) instead of local defects (singular intersite disorder).
The distinction between local and extended defects are important especially in theoretical modeling of disordered Kitaev magnets~\cite{kao_vacancy-induced_2020}.

In the inset of Fig.~\ref{fig:TEM}(a), a crystallographic model is overlaid on the magnified image to identify the Ag, Ir, Li, and O atoms as blue, yellow, pink, and red spheres, respectively (only the Ag and Ir atoms are clearly visible).
The arrows indicate where the unwanted Ag atoms (blue) are inserted within the Ir layer (yellow).
In contrast, the HAADF-STEM image from the clean sample S1 in Fig.~\ref{fig:TEM}(b) shows pristine honeycomb layers free from silver inclusions.

We present the electron diffraction (ED) patterns for S1 and S2 in the top insets of Fig.~\ref{fig:TEM}(a,b).
The streaking in ED patterns is due to the stacking faults in the form of angular twist between the adjacent layers as shown in other honeycomb materials~\cite{abramchuk_crystal_2018}.
Upon careful inspection, the ED pattern of sample S1 reveals less streaking than S2.
This is consistent with the synthesis of sample S1 from a precursor \LIO\ with less stacking faults as explained in Appendix~\ref{app:synthesis} (Fig.~\ref{fig:LIO}). 
We show in Appendix~\ref{app:TEM} (Fig.~\ref{fig:TEM2}) that \ALIO\ has more stacking faults than its parent compounds \LIO.
It is likely that in the absence of such stacking faults, the initial spin freezing at $T_F$ could turn into a long-range order, i.e. $T_F=T_N$~\cite{rousochatzakis_quantum_2019}.

%%%%%%%%%%%%%%%%%%%%%%%%%%%%%%%%%%%%%%%%%%%%%%%
\section{\label{sec:conclusion}CONCLUSION}
%%%%%%%%%%%%%%%%%%%%%%%%%%%%%%%%%%%%%%%%%%%%%%%
By improving the sample quality, we have revealed signatures of a long-range incommensurate order in \ALIO.
A broad peak in the magnetic susceptibility and heat capacity at $T_F=14$~K marks the onset of spin freezing.
Such a peak is absent in the disordered sample, which hinders the recognition of a long-range order in \ALIO.
In $\mu$SR, a fast decay of muon depolarization below $T_F$ confirms the onset of static magnetism, and the appearance of oscillations below $T_N$ confirm the long-range order.
The oscillation patterns at low temperatures fit to a Bessel function, consistent with incommensurate ordering.
Such an incommensurate order has been confirmed in \LIO\ from both $\mu$SR and neutron scattering~\cite{choi_spin_2019}.
Thus, we conclude that the change of interlayer atoms does not affect the magnetic ordering in \LIO, i.e. the magnetic ground states of \ALIO\ and \LIO\ are the same.
Our HAADF-STEM images confirm a moderate level of extended defects (silver inclusion) in the disordered \ALIO\ sample made from a lower quality \LIO. 
In the disordered sample, the Ag atoms enter the honeycomb layer and disrupt the long-range magnetic order.
This effect must be distinguished from the lack of magnetic ordering due to long-range entanglement in a quantum spin liquid.

\section*{ACKNOWLEDGMENTS}
The first two authors have contributed equally to this work.
We thank R.~Valenti and N.~B.~Perkins for fruitful discussions, and H.~Leutkens, T.~Shiroka and C.~Baines for their technical assistance with the $\mu$SR experiments.
The work at Boston College was supported by the National Science Foundation under award No. DMR--1708929.
This work is based on experiments performed at the Swiss Muon Source S$\mu$S at the Paul Scherrer Institute, Villigen, Switzerland, and at the ISIS Pulsed Muon Source, which is supported by the UK Science and Technology Facilities Council.

\appendix
%%%%%%%%%%%%%%%%%%%%%%%%%%%%%%%%%%%%%%%%%%%%%%%
\section{\label{app:synthesis}Synthesis details.}
%%%%%%%%%%%%%%%%%%%%%%%%%%%%%%%%%%%%%%%%%%%%%%%
\begin{figure}
\includegraphics[width=0.46\textwidth]{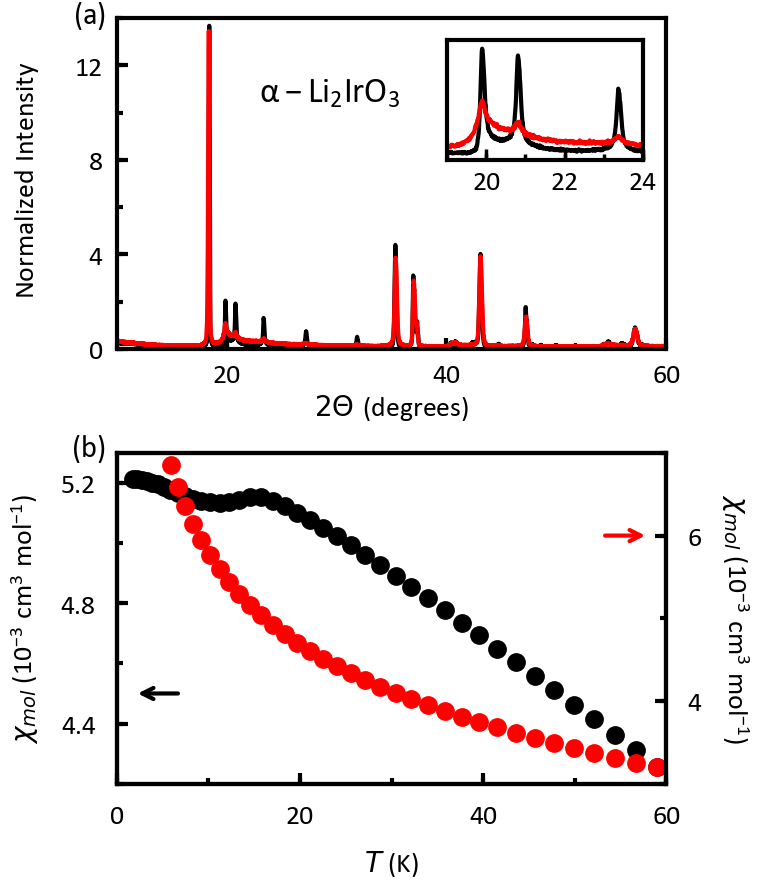}
\caption{\label{fig:LIO}
(a) X-ray patterns of two \LIO\ precursors used in the synthesis of clean (black) and disordered (red) \ALIO. 
The region of honeycomb peaks is magnified in the inset.
(b) Temperature dependence of the DC magnetic susceptibility in the two \LIO\ precursors, with the same color code as in panel (a).}
\end{figure}
%%%%%%%%%%%%%%%%%%%%%%%%%%%%%%%%%%%%%%%%%%%%%%%
The important difference between the two \ALIO\ samples, S1 and S2, is in the \LIO\ precursor used in their synthesis. 
Figure~\ref{fig:LIO}(a) compares the x-ray patterns between two \LIO\ precursors, shown in black and red, used for the synthesis of samples S1 (clean) and S2 (disordered), respectively. 
The region between 19$^{\circ}$ to 24$^{\circ}$ gives information about the quality of honeycomb ordering in \LIO. 
The black x-ray pattern with sharp and well-separated peaks indicates better honeycomb ordering and less stacking faults than the red x-ray pattern. 
A similar level of disorder carries over to the \ALIO\ produced from these precursors.
We also reveal the effect of disorder on the magnetic behavior of \LIO\ in Fig.~\ref{fig:LIO}(b), by plotting DC susceptibility of both \LIO\ samples as a function of temperature below 60~K.
The red curve does not show any signs of magnetic ordering while the black curve shows a peak at the AFM transition at 15~K.

% To synthesize \LIO, a well ground mixture of lithium carbonate (Li$_{2}$CO$_{3}$, Alfa Aesar, 99.998$\%$, dried at 120 $^{\circ}$C for 24 h) and iridium oxide (IrO$_{2}$, Alfa Aesar, 99$\%$) was prepared with the mole ratio of 1.2:1.
% %
% The mixture with total mass of 350~mg was pressed into a 6~mm diameter pellet and placed in a covered alumina crucible. 
% % 
% The crucible was heated at 5~\C/min to 900~\C\ and held at that temperature for 16~h, then cooled at 5~\C/min to 600~\C, and quenched in the antechamber of the glovebox (H$_{2}$O and O$_{2}$ contents $<$ 1 ppm). 
% %
% To get rid of unreacted IrO$_2$ and improve the stacking faults, the resulting sample was ground and annealed several times at 1000/1015~\C\ for 24/48~h. 
% %
% The topotactic cation-exchange reaction was then performed according to 
% \begin{equation}
% \mathrm{2Li_2IrO_3 + 3AgNO_3 \rightarrow Ag_3LiIr_2O_6 + 3LiNO_3}
% \end{equation}
% %
% from a mixture of \LIO\ and Silver nitrate (AgNO$_{3}$, Alfa Aeser, 99.9+$\%$).
% %
% The mixture with total mass of 300~mg was pelletized, placed in a covered alumina crucible, and sealed
% under a partial argon pressure of 10 inHg inside a quartz tube. 
% %
% The tube was heated at 1~\C/min to 350~\C\ and kept at that temperature for 7 days, then cooled to room temperature at the same rate. 
% %
% To remove the excess AgNO$_{3}$, the sample was washed three times with deionized water and dried at room temperature under vacuum for 1~h.

%%%%%%%%%%%%%%%%%%%%%%%%%%%%%%%%%%%%%%%%%%%%%%%
\section{\label{app:zfc}Splitting between ZFC and FC data.}
%%%%%%%%%%%%%%%%%%%%%%%%%%%%%%%%%%%%%%%%%%%%%%%
\begin{figure}
\includegraphics[width=0.46\textwidth]{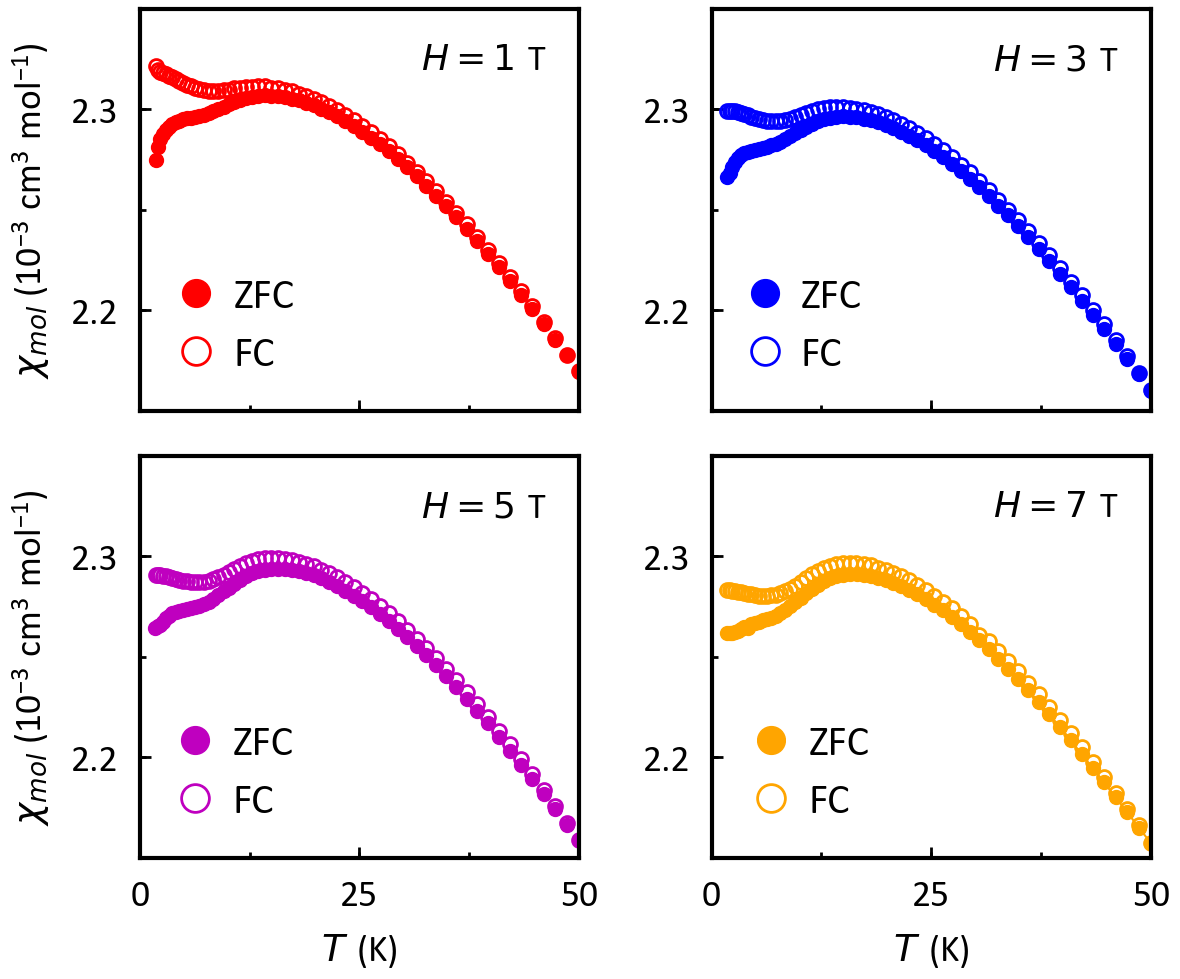}
\caption{\label{fig:ZFCFC}
The splitting between ZFC (full circles) and FC (open circles) susceptibility curves at 1, 3, 5, and 7~T.}
\end{figure}
%%%%%%%%%%%%%%%%%%%%%%%%%%%%%%%%%%%%%%%%%%%%%%%
In Fig.~\ref{fig:ZFCFC}, we show the splitting between ZFC and FC susceptibility at several fields.
Note that the splitting persists to high fields, confirming a static spin freezing~\cite{raju_magnetic-susceptibility_1992,mydosh_spin_2015} at $T_F$, as noted in the main text.

\section{\label{app:muH}$\mu$SR data under longitudinal field.}
%%%%%%%%%%%%%%%%%%%%%%%%%%%%%%%%%%%%%%%%%%%%%%%
\begin{figure}
\includegraphics[width=0.46\textwidth]{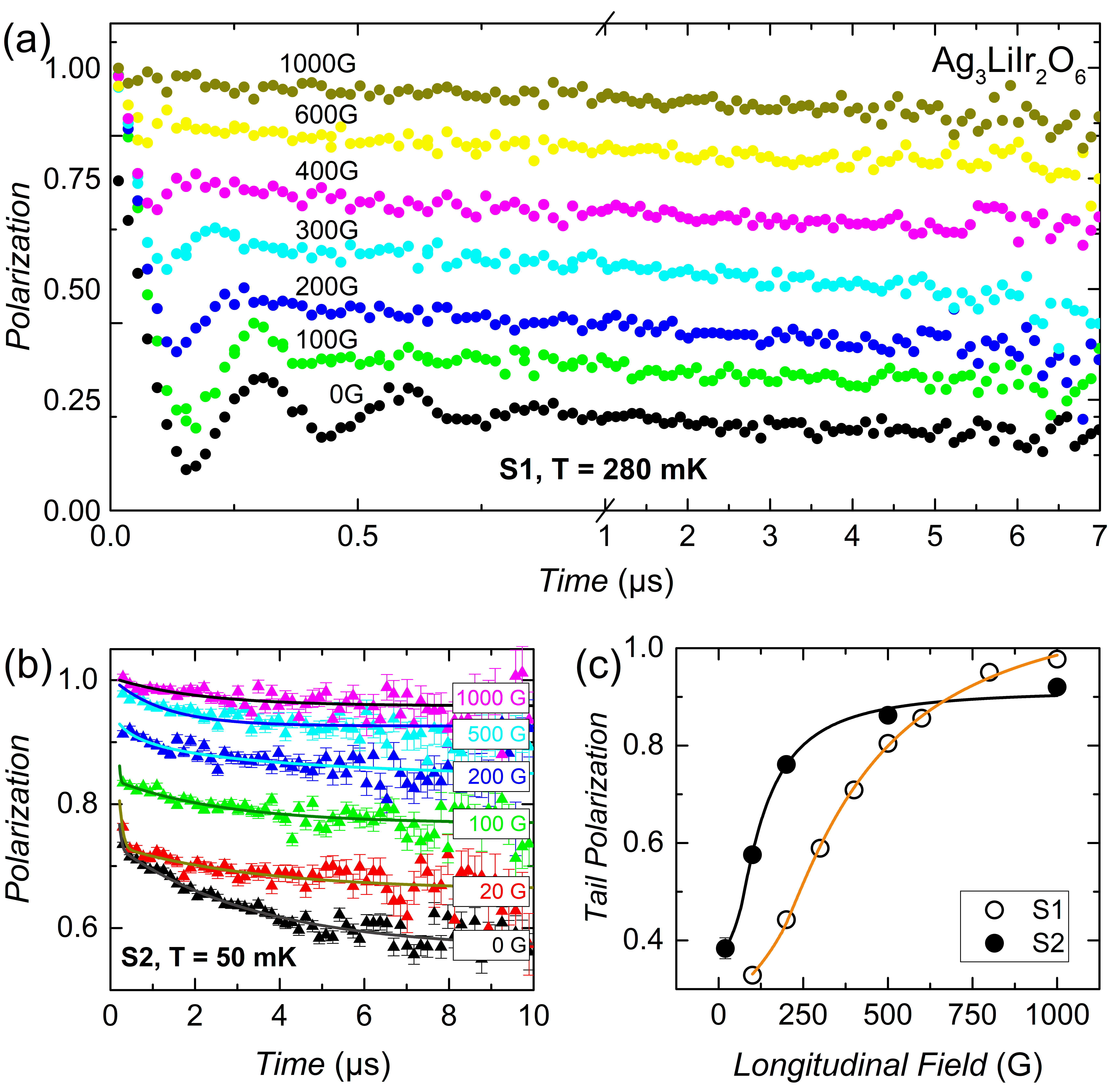}
\caption{\label{fig:LF}
(a) Polarization scans in the clean sample S1 at 0.28~K under different longitudinal fields (LF) from 0 to 1000~G.
The time axis is expanded for $t<1~\mu$s to reveal the oscillations.
(b) Polarization scans in the disordered sample S2 at 0.05~K under LF from 0 to 1000~G.
The data in panels (a) and (b) were collected at the PSI and ISIS facilities, respectively.
(c) By analyzing the recovery of the initial asymmetry with increasing field, we estimate $B_{\textrm{int}}=263$~G in S1 and $113$~G in S2.
Solid lines are guides to the eye.
}
\end{figure}
%%%%%%%%%%%%%%%%%%%%%%%%%%%%%%%%%%%%%%%%%%%%%%%
In the main text, we derived $B_{\mathrm{max}}=269$~G in sample S1 at 0.28~K by fitting the zero-field (ZF) $\mu$SR data to a Bessel function (Eq.~\ref{eq:bessel}).
As a consistency check, here we estimate the internal field $B_{\mathrm{int}}$ by analyzing the longitudinal field (LF) scans at 0.28~K as shown in Fig.~\ref{fig:LF}(a). 
The initial polarization is fully recovered by 1000~G, so the internal field $B_{\mathrm{int}}$ must be much smaller than this value. 
A detailed analysis~\cite{pratt_field_2007} shows that the midpoint of the polarization recovery occurs at a field value close to $B/B_{\mathrm{int}}=4/3$. 
Figure~\ref{fig:LF}(c) shows that the midpoint of recovery in S1 is at 350~G, yielding an internal field $B_{\textrm{int}}=263$~G, in good agreement with the $B_{\mathrm{max}}=269$~G obtained from our Bessel function fit to Eq.~\ref{eq:bessel}.
We have also collected LF scans from the disordered sample S2 at 0.05~K as shown in Fig.~\ref{fig:LF}(b).
The midpoint of recovery in S2 occurs at 150~G in Fig.~\ref{fig:LF}(c), yielding an internal field $B_{\textrm{int}}=113$~G which is smaller than in sample S1.
A smaller internal field may result from a range of muon stopping sites in the disordered sample.
Since $\mu$SR is a local probe, we do not expect a major change in the local field near Ir$^{4+}$ sites below $T_N$, but it is likely that muons probe a range of stopping sites with slightly different chemical environment due to various levels of Ag inclusion across the sample.
This explains the slow depolarization of muons inside S2 at 10~K in Fig.~\ref{fig:OP}, and the different polarization recovery between S2 and S1 in Fig.~\ref{fig:LF}(c).
As noted in the main text, it is not possible to fit the ZF data in sample S2 to a Bessel function (Eq.~\ref{eq:bessel}) because the oscillations are not discernible in the disordered sample. 
Thus, the LF analysis is the only way of estimating the local internal field in S2.

%%%%%%%%%%%%%%%%%%%%%%%%%%%%%%%%%%%%%%%%%%%%%%%
\section{\label{app:TEM}TEM analysis of the stacking faults.}
%%%%%%%%%%%%%%%%%%%%%%%%%%%%%%%%%%%%%%%%%%%%%%%
\begin{figure}
\includegraphics[width=0.5\textwidth]{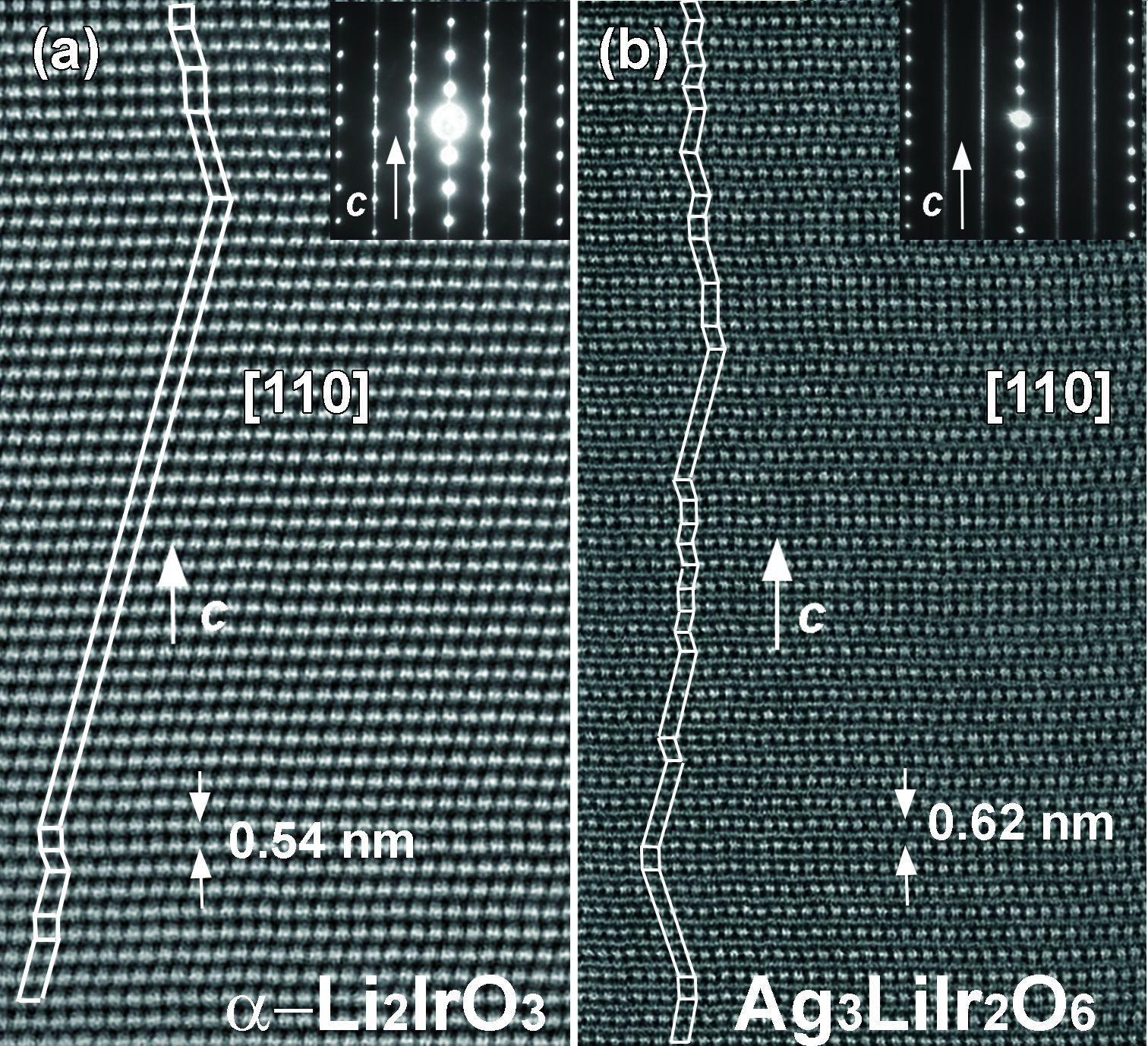}
\caption{\label{fig:TEM2}
HAADF-TEM images from (a) \LIO\ and (b) \ALIO\ (S1).
A clean sample is used for each material.
The magnetization and X-ray data for the \LIO\ sample are presented in Fig.~\ref{fig:LIO} (black data).
The magnetization data for the \ALIO\ are presented in the main text (sample S1).
The images show an abundance of stacking faults in \ALIO\ unlike \LIO, due to the weaker interlayer coupling in the former.
The ED patterns are presented as insets and reveal less streaking in \LIO\ due to less stacking faults compared to \ALIO.}
\end{figure}
%%%%%%%%%%%%%%%%%%%%%%%%%%%%%%%%%%%%%%%%%%%%%%%
Our discussion of the structural disorder in the main text is focused on the Ag inclusion within the honeycomb layers of \ALIO\ (Fig.~\ref{fig:TEM}).
Here we point out that both the clean (S1) and disordered (S2) samples of \ALIO\ also suffer from the stacking faults, similar to other layered honeycomb materials such as Cu$_3$LiSn$_2$O$_6$~\cite{abramchuk_crystal_2018}.
%However, the impact of stacking faults on the magnetic odering is much less than the effect of silver inclusion.
%
Figure~\ref{fig:TEM2} compares HAADF-TEM images between a clean sample of \LIO\ and a clean sample of \ALIO\ (S1).
There is no intersite disorder in either image, but \ALIO\ exhibits much more stacking faults than its parent compound \LIO.
It has been demonstrated in a prior study of Cu$_3$LiSn$_2$O$_6$ that the stacking faults result from a twisting between the adjacent honeycomb layers, due to the weak O-Cu-O dumbbell bonds between the layers~\cite{abramchuk_crystal_2018}.
A similar mechanism is at work in \ALIO, where the weak O-Ag-O dumbbell bonds lead to the twisting between the layers and produce the zig-zag stacking pattern observed in Fig.~\ref{fig:TEM2}(b).
Despite the considerable amount of stacking faults in sample S1 (Fig.~\ref{fig:TEM2}(b)), it still shows clear signatures of long-range order as explained in the main text.
In fact, the incommensurate order is similar between \LIO\ and \ALIO\ based on our $\mu$SR data and the published results in Reference~\cite{choi_spin_2019}.
Thus, we conclude that the magnetic interactions within the honeycomb layers are not affected by the interlayer bonds; however, they are disrupted by the extended defects in form of silver inclusion within the honeycomb layers.

%%%%%%%%%%%%%%%%%%%%%%%%%%%%%%%%%%%%%%%%%%%%%%%
\bibliography{Bahrami_12nov2020}% Produces the bibliography via BibTeX.

%merlin.mbs apsrev4-1.bst 2010-07-25 4.21a (PWD, AO, DPC) hacked
%Control: key (0)
%Control: author (8) initials jnrlst
%Control: editor formatted (1) identically to author
%Control: production of article title (-1) disabled
%Control: page (0) single
%Control: year (1) truncated
%Control: production of eprint (0) enabled
\begin{thebibliography}{32}%
\makeatletter
\providecommand \@ifxundefined [1]{%
 \@ifx{#1\undefined}
}%
\providecommand \@ifnum [1]{%
 \ifnum #1\expandafter \@firstoftwo
 \else \expandafter \@secondoftwo
 \fi
}%
\providecommand \@ifx [1]{%
 \ifx #1\expandafter \@firstoftwo
 \else \expandafter \@secondoftwo
 \fi
}%
\providecommand \natexlab [1]{#1}%
\providecommand \enquote  [1]{``#1''}%
\providecommand \bibnamefont  [1]{#1}%
\providecommand \bibfnamefont [1]{#1}%
\providecommand \citenamefont [1]{#1}%
\providecommand \href@noop [0]{\@secondoftwo}%
\providecommand \href [0]{\begingroup \@sanitize@url \@href}%
\providecommand \@href[1]{\@@startlink{#1}\@@href}%
\providecommand \@@href[1]{\endgroup#1\@@endlink}%
\providecommand \@sanitize@url [0]{\catcode `\\12\catcode `\$12\catcode
  `\&12\catcode `\#12\catcode `\^12\catcode `\_12\catcode `\%12\relax}%
\providecommand \@@startlink[1]{}%
\providecommand \@@endlink[0]{}%
\providecommand \url  [0]{\begingroup\@sanitize@url \@url }%
\providecommand \@url [1]{\endgroup\@href {#1}{\urlprefix }}%
\providecommand \urlprefix  [0]{URL }%
\providecommand \Eprint [0]{\href }%
\providecommand \doibase [0]{http://dx.doi.org/}%
\providecommand \selectlanguage [0]{\@gobble}%
\providecommand \bibinfo  [0]{\@secondoftwo}%
\providecommand \bibfield  [0]{\@secondoftwo}%
\providecommand \translation [1]{[#1]}%
\providecommand \BibitemOpen [0]{}%
\providecommand \bibitemStop [0]{}%
\providecommand \bibitemNoStop [0]{.\EOS\space}%
\providecommand \EOS [0]{\spacefactor3000\relax}%
\providecommand \BibitemShut  [1]{\csname bibitem#1\endcsname}%
\let\auto@bib@innerbib\@empty
%</preamble>
\bibitem [{\citenamefont {Broholm}\ \emph {et~al.}(2020)\citenamefont
  {Broholm}, \citenamefont {Cava}, \citenamefont {Kivelson}, \citenamefont
  {Nocera}, \citenamefont {Norman},\ and\ \citenamefont
  {Senthil}}]{broholm_quantum_2020}%
  \BibitemOpen
  \bibfield  {author} {\bibinfo {author} {\bibfnamefont {C.}~\bibnamefont
  {Broholm}}, \bibinfo {author} {\bibfnamefont {R.~J.}\ \bibnamefont {Cava}},
  \bibinfo {author} {\bibfnamefont {S.~A.}\ \bibnamefont {Kivelson}}, \bibinfo
  {author} {\bibfnamefont {D.~G.}\ \bibnamefont {Nocera}}, \bibinfo {author}
  {\bibfnamefont {M.~R.}\ \bibnamefont {Norman}}, \ and\ \bibinfo {author}
  {\bibfnamefont {T.}~\bibnamefont {Senthil}},\ }\href {\doibase
  10.1126/science.aay0668} {\bibfield  {journal} {\bibinfo  {journal}
  {Science}\ }\textbf {\bibinfo {volume} {367}} (\bibinfo {year} {2020}),\
  10.1126/science.aay0668}\BibitemShut {NoStop}%
\bibitem [{\citenamefont {Knolle}\ and\ \citenamefont
  {Moessner}(2019)}]{knolle_field_2019}%
  \BibitemOpen
  \bibfield  {author} {\bibinfo {author} {\bibfnamefont {J.}~\bibnamefont
  {Knolle}}\ and\ \bibinfo {author} {\bibfnamefont {R.}~\bibnamefont
  {Moessner}},\ }\href {\doibase 10.1146/annurev-conmatphys-031218-013401}
  {\bibfield  {journal} {\bibinfo  {journal} {Annual Review of Condensed Matter
  Physics}\ }\textbf {\bibinfo {volume} {10}},\ \bibinfo {pages} {451}
  (\bibinfo {year} {2019})},\ \bibinfo {note} {\_eprint:
  https://doi.org/10.1146/annurev-conmatphys-031218-013401}\BibitemShut
  {NoStop}%
\bibitem [{\citenamefont {Savary}\ and\ \citenamefont
  {Balents}(2016)}]{savary_quantum_2016}%
  \BibitemOpen
  \bibfield  {author} {\bibinfo {author} {\bibfnamefont {L.}~\bibnamefont
  {Savary}}\ and\ \bibinfo {author} {\bibfnamefont {L.}~\bibnamefont
  {Balents}},\ }\href {\doibase 10.1088/0034-4885/80/1/016502} {\bibfield
  {journal} {\bibinfo  {journal} {Reports on Progress in Physics}\ }\textbf
  {\bibinfo {volume} {80}},\ \bibinfo {pages} {016502} (\bibinfo {year}
  {2016})}\BibitemShut {NoStop}%
\bibitem [{\citenamefont {Kitaev}(2006)}]{kitaev_anyons_2006}%
  \BibitemOpen
  \bibfield  {author} {\bibinfo {author} {\bibfnamefont {A.}~\bibnamefont
  {Kitaev}},\ }\href {\doibase 10.1016/j.aop.2005.10.005} {\bibfield  {journal}
  {\bibinfo  {journal} {Annals of Physics}\ }\bibinfo {series} {January
  {Special} {Issue}},\ \textbf {\bibinfo {volume} {321}},\ \bibinfo {pages} {2}
  (\bibinfo {year} {2006})}\BibitemShut {NoStop}%
\bibitem [{\citenamefont {Jackeli}\ and\ \citenamefont
  {Khaliullin}(2009)}]{jackeli_mott_2009}%
  \BibitemOpen
  \bibfield  {author} {\bibinfo {author} {\bibfnamefont {G.}~\bibnamefont
  {Jackeli}}\ and\ \bibinfo {author} {\bibfnamefont {G.}~\bibnamefont
  {Khaliullin}},\ }\href {\doibase 10.1103/PhysRevLett.102.017205} {\bibfield
  {journal} {\bibinfo  {journal} {Physical Review Letters}\ }\textbf {\bibinfo
  {volume} {102}},\ \bibinfo {pages} {017205} (\bibinfo {year}
  {2009})}\BibitemShut {NoStop}%
\bibitem [{\citenamefont {Singh}\ \emph {et~al.}(2012)\citenamefont {Singh},
  \citenamefont {Manni}, \citenamefont {Reuther}, \citenamefont {Berlijn},
  \citenamefont {Thomale}, \citenamefont {Ku}, \citenamefont {Trebst},\ and\
  \citenamefont {Gegenwart}}]{singh_relevance_2012}%
  \BibitemOpen
  \bibfield  {author} {\bibinfo {author} {\bibfnamefont {Y.}~\bibnamefont
  {Singh}}, \bibinfo {author} {\bibfnamefont {S.}~\bibnamefont {Manni}},
  \bibinfo {author} {\bibfnamefont {J.}~\bibnamefont {Reuther}}, \bibinfo
  {author} {\bibfnamefont {T.}~\bibnamefont {Berlijn}}, \bibinfo {author}
  {\bibfnamefont {R.}~\bibnamefont {Thomale}}, \bibinfo {author} {\bibfnamefont
  {W.}~\bibnamefont {Ku}}, \bibinfo {author} {\bibfnamefont {S.}~\bibnamefont
  {Trebst}}, \ and\ \bibinfo {author} {\bibfnamefont {P.}~\bibnamefont
  {Gegenwart}},\ }\href {\doibase 10.1103/PhysRevLett.108.127203} {\bibfield
  {journal} {\bibinfo  {journal} {Physical Review Letters}\ }\textbf {\bibinfo
  {volume} {108}},\ \bibinfo {pages} {127203} (\bibinfo {year}
  {2012})}\BibitemShut {NoStop}%
\bibitem [{\citenamefont {Takagi}\ \emph {et~al.}(2019)\citenamefont {Takagi},
  \citenamefont {Takayama}, \citenamefont {Jackeli}, \citenamefont
  {Khaliullin},\ and\ \citenamefont {Nagler}}]{takagi_concept_2019}%
  \BibitemOpen
  \bibfield  {author} {\bibinfo {author} {\bibfnamefont {H.}~\bibnamefont
  {Takagi}}, \bibinfo {author} {\bibfnamefont {T.}~\bibnamefont {Takayama}},
  \bibinfo {author} {\bibfnamefont {G.}~\bibnamefont {Jackeli}}, \bibinfo
  {author} {\bibfnamefont {G.}~\bibnamefont {Khaliullin}}, \ and\ \bibinfo
  {author} {\bibfnamefont {S.~E.}\ \bibnamefont {Nagler}},\ }\href {\doibase
  10.1038/s42254-019-0038-2} {\bibfield  {journal} {\bibinfo  {journal} {Nature
  Reviews Physics}\ }\textbf {\bibinfo {volume} {1}},\ \bibinfo {pages} {264}
  (\bibinfo {year} {2019})}\BibitemShut {NoStop}%
\bibitem [{\citenamefont {Plumb}\ \emph {et~al.}(2014)\citenamefont {Plumb},
  \citenamefont {Clancy}, \citenamefont {Sandilands}, \citenamefont {Shankar},
  \citenamefont {Hu}, \citenamefont {Burch}, \citenamefont {Kee},\ and\
  \citenamefont {Kim}}]{plumb_rucl_3_2014}%
  \BibitemOpen
  \bibfield  {author} {\bibinfo {author} {\bibfnamefont {K.~W.}\ \bibnamefont
  {Plumb}}, \bibinfo {author} {\bibfnamefont {J.~P.}\ \bibnamefont {Clancy}},
  \bibinfo {author} {\bibfnamefont {L.~J.}\ \bibnamefont {Sandilands}},
  \bibinfo {author} {\bibfnamefont {V.~V.}\ \bibnamefont {Shankar}}, \bibinfo
  {author} {\bibfnamefont {Y.~F.}\ \bibnamefont {Hu}}, \bibinfo {author}
  {\bibfnamefont {K.~S.}\ \bibnamefont {Burch}}, \bibinfo {author}
  {\bibfnamefont {H.-Y.}\ \bibnamefont {Kee}}, \ and\ \bibinfo {author}
  {\bibfnamefont {Y.-J.}\ \bibnamefont {Kim}},\ }\href {\doibase
  10.1103/PhysRevB.90.041112} {\bibfield  {journal} {\bibinfo  {journal}
  {Physical Review B}\ }\textbf {\bibinfo {volume} {90}},\ \bibinfo {pages}
  {041112} (\bibinfo {year} {2014})}\BibitemShut {NoStop}%
\bibitem [{\citenamefont {Nasu}\ \emph {et~al.}(2016)\citenamefont {Nasu},
  \citenamefont {Knolle}, \citenamefont {Kovrizhin}, \citenamefont {Motome},\
  and\ \citenamefont {Moessner}}]{nasu_fermionic_2016}%
  \BibitemOpen
  \bibfield  {author} {\bibinfo {author} {\bibfnamefont {J.}~\bibnamefont
  {Nasu}}, \bibinfo {author} {\bibfnamefont {J.}~\bibnamefont {Knolle}},
  \bibinfo {author} {\bibfnamefont {D.~L.}\ \bibnamefont {Kovrizhin}}, \bibinfo
  {author} {\bibfnamefont {Y.}~\bibnamefont {Motome}}, \ and\ \bibinfo {author}
  {\bibfnamefont {R.}~\bibnamefont {Moessner}},\ }\href {\doibase
  10.1038/nphys3809} {\bibfield  {journal} {\bibinfo  {journal} {Nature
  Physics}\ }\textbf {\bibinfo {volume} {12}},\ \bibinfo {pages} {912}
  (\bibinfo {year} {2016})}\BibitemShut {NoStop}%
\bibitem [{\citenamefont {Wang}\ \emph {et~al.}(2020)\citenamefont {Wang},
  \citenamefont {Osterhoudt}, \citenamefont {Tian}, \citenamefont
  {Lampen-Kelley}, \citenamefont {Banerjee}, \citenamefont {Goldstein},
  \citenamefont {Yan}, \citenamefont {Knolle}, \citenamefont {Ji},
  \citenamefont {Cava}, \citenamefont {Nasu}, \citenamefont {Motome},
  \citenamefont {Nagler}, \citenamefont {Mandrus},\ and\ \citenamefont
  {Burch}}]{wang_range_2020}%
  \BibitemOpen
  \bibfield  {author} {\bibinfo {author} {\bibfnamefont {Y.}~\bibnamefont
  {Wang}}, \bibinfo {author} {\bibfnamefont {G.~B.}\ \bibnamefont
  {Osterhoudt}}, \bibinfo {author} {\bibfnamefont {Y.}~\bibnamefont {Tian}},
  \bibinfo {author} {\bibfnamefont {P.}~\bibnamefont {Lampen-Kelley}}, \bibinfo
  {author} {\bibfnamefont {A.}~\bibnamefont {Banerjee}}, \bibinfo {author}
  {\bibfnamefont {T.}~\bibnamefont {Goldstein}}, \bibinfo {author}
  {\bibfnamefont {J.}~\bibnamefont {Yan}}, \bibinfo {author} {\bibfnamefont
  {J.}~\bibnamefont {Knolle}}, \bibinfo {author} {\bibfnamefont
  {H.}~\bibnamefont {Ji}}, \bibinfo {author} {\bibfnamefont {R.~J.}\
  \bibnamefont {Cava}}, \bibinfo {author} {\bibfnamefont {J.}~\bibnamefont
  {Nasu}}, \bibinfo {author} {\bibfnamefont {Y.}~\bibnamefont {Motome}},
  \bibinfo {author} {\bibfnamefont {S.~E.}\ \bibnamefont {Nagler}}, \bibinfo
  {author} {\bibfnamefont {D.}~\bibnamefont {Mandrus}}, \ and\ \bibinfo
  {author} {\bibfnamefont {K.~S.}\ \bibnamefont {Burch}},\ }\href {\doibase
  10.1038/s41535-020-0216-6} {\bibfield  {journal} {\bibinfo  {journal} {npj
  Quantum Materials}\ }\textbf {\bibinfo {volume} {5}},\ \bibinfo {pages} {1}
  (\bibinfo {year} {2020})},\ \bibinfo {note} {number: 1 Publisher: Nature
  Publishing Group}\BibitemShut {NoStop}%
\bibitem [{\citenamefont {Kitagawa}\ \emph {et~al.}(2018)\citenamefont
  {Kitagawa}, \citenamefont {Takayama}, \citenamefont {Matsumoto},
  \citenamefont {Kato}, \citenamefont {Takano}, \citenamefont {Kishimoto},
  \citenamefont {Bette}, \citenamefont {Dinnebier}, \citenamefont {Jackeli},\
  and\ \citenamefont {Takagi}}]{kitagawa_spinorbital-entangled_2018}%
  \BibitemOpen
  \bibfield  {author} {\bibinfo {author} {\bibfnamefont {K.}~\bibnamefont
  {Kitagawa}}, \bibinfo {author} {\bibfnamefont {T.}~\bibnamefont {Takayama}},
  \bibinfo {author} {\bibfnamefont {Y.}~\bibnamefont {Matsumoto}}, \bibinfo
  {author} {\bibfnamefont {A.}~\bibnamefont {Kato}}, \bibinfo {author}
  {\bibfnamefont {R.}~\bibnamefont {Takano}}, \bibinfo {author} {\bibfnamefont
  {Y.}~\bibnamefont {Kishimoto}}, \bibinfo {author} {\bibfnamefont
  {S.}~\bibnamefont {Bette}}, \bibinfo {author} {\bibfnamefont
  {R.}~\bibnamefont {Dinnebier}}, \bibinfo {author} {\bibfnamefont
  {G.}~\bibnamefont {Jackeli}}, \ and\ \bibinfo {author} {\bibfnamefont
  {H.}~\bibnamefont {Takagi}},\ }\href {\doibase 10.1038/nature25482}
  {\bibfield  {journal} {\bibinfo  {journal} {Nature}\ }\textbf {\bibinfo
  {volume} {554}},\ \bibinfo {pages} {341} (\bibinfo {year}
  {2018})}\BibitemShut {NoStop}%
\bibitem [{\citenamefont {Roudebush}\ \emph {et~al.}(2016)\citenamefont
  {Roudebush}, \citenamefont {Ross},\ and\ \citenamefont
  {Cava}}]{roudebush_iridium_2016}%
  \BibitemOpen
  \bibfield  {author} {\bibinfo {author} {\bibfnamefont {J.~H.}\ \bibnamefont
  {Roudebush}}, \bibinfo {author} {\bibfnamefont {K.~A.}\ \bibnamefont {Ross}},
  \ and\ \bibinfo {author} {\bibfnamefont {R.~J.}\ \bibnamefont {Cava}},\
  }\href {\doibase 10.1039/C6DT00798H} {\bibfield  {journal} {\bibinfo
  {journal} {Dalton Transactions}\ }\textbf {\bibinfo {volume} {45}},\ \bibinfo
  {pages} {8783} (\bibinfo {year} {2016})}\BibitemShut {NoStop}%
\bibitem [{\citenamefont {Abramchuk}\ \emph {et~al.}(2017)\citenamefont
  {Abramchuk}, \citenamefont {Ozsoy-Keskinbora}, \citenamefont {Krizan},
  \citenamefont {Metz}, \citenamefont {Bell},\ and\ \citenamefont
  {Tafti}}]{abramchuk_cu2iro3_2017}%
  \BibitemOpen
  \bibfield  {author} {\bibinfo {author} {\bibfnamefont {M.}~\bibnamefont
  {Abramchuk}}, \bibinfo {author} {\bibfnamefont {C.}~\bibnamefont
  {Ozsoy-Keskinbora}}, \bibinfo {author} {\bibfnamefont {J.~W.}\ \bibnamefont
  {Krizan}}, \bibinfo {author} {\bibfnamefont {K.~R.}\ \bibnamefont {Metz}},
  \bibinfo {author} {\bibfnamefont {D.~C.}\ \bibnamefont {Bell}}, \ and\
  \bibinfo {author} {\bibfnamefont {F.}~\bibnamefont {Tafti}},\ }\href
  {\doibase 10.1021/jacs.7b06911} {\bibfield  {journal} {\bibinfo  {journal}
  {Journal of the American Chemical Society}\ }\textbf {\bibinfo {volume}
  {139}},\ \bibinfo {pages} {15371} (\bibinfo {year} {2017})}\BibitemShut
  {NoStop}%
\bibitem [{\citenamefont {Bahrami}\ \emph {et~al.}(2019)\citenamefont
  {Bahrami}, \citenamefont {Lafargue-Dit-Hauret}, \citenamefont {Lebedev},
  \citenamefont {Movshovich}, \citenamefont {Yang}, \citenamefont {Broido},
  \citenamefont {Rocquefelte},\ and\ \citenamefont
  {Tafti}}]{bahrami_thermodynamic_2019}%
  \BibitemOpen
  \bibfield  {author} {\bibinfo {author} {\bibfnamefont {F.}~\bibnamefont
  {Bahrami}}, \bibinfo {author} {\bibfnamefont {W.}~\bibnamefont
  {Lafargue-Dit-Hauret}}, \bibinfo {author} {\bibfnamefont {O.~I.}\
  \bibnamefont {Lebedev}}, \bibinfo {author} {\bibfnamefont {R.}~\bibnamefont
  {Movshovich}}, \bibinfo {author} {\bibfnamefont {H.-Y.}\ \bibnamefont
  {Yang}}, \bibinfo {author} {\bibfnamefont {D.}~\bibnamefont {Broido}},
  \bibinfo {author} {\bibfnamefont {X.}~\bibnamefont {Rocquefelte}}, \ and\
  \bibinfo {author} {\bibfnamefont {F.}~\bibnamefont {Tafti}},\ }\href
  {\doibase 10.1103/PhysRevLett.123.237203} {\bibfield  {journal} {\bibinfo
  {journal} {Physical Review Letters}\ }\textbf {\bibinfo {volume} {123}},\
  \bibinfo {pages} {237203} (\bibinfo {year} {2019})}\BibitemShut {NoStop}%
\bibitem [{\citenamefont {Kimchi}\ \emph {et~al.}(2018)\citenamefont {Kimchi},
  \citenamefont {Sheckelton}, \citenamefont {McQueen},\ and\ \citenamefont
  {Lee}}]{kimchi_scaling_2018}%
  \BibitemOpen
  \bibfield  {author} {\bibinfo {author} {\bibfnamefont {I.}~\bibnamefont
  {Kimchi}}, \bibinfo {author} {\bibfnamefont {J.~P.}\ \bibnamefont
  {Sheckelton}}, \bibinfo {author} {\bibfnamefont {T.~M.}\ \bibnamefont
  {McQueen}}, \ and\ \bibinfo {author} {\bibfnamefont {P.~A.}\ \bibnamefont
  {Lee}},\ }\href {\doibase 10.1038/s41467-018-06800-2} {\bibfield  {journal}
  {\bibinfo  {journal} {Nature Communications}\ }\textbf {\bibinfo {volume}
  {9}},\ \bibinfo {pages} {4367} (\bibinfo {year} {2018})}\BibitemShut
  {NoStop}%
\bibitem [{\citenamefont {Knolle}\ \emph {et~al.}(2019)\citenamefont {Knolle},
  \citenamefont {Moessner},\ and\ \citenamefont
  {Perkins}}]{knolle_bond-disordered_2019}%
  \BibitemOpen
  \bibfield  {author} {\bibinfo {author} {\bibfnamefont {J.}~\bibnamefont
  {Knolle}}, \bibinfo {author} {\bibfnamefont {R.}~\bibnamefont {Moessner}}, \
  and\ \bibinfo {author} {\bibfnamefont {N.~B.}\ \bibnamefont {Perkins}},\
  }\href {\doibase 10.1103/PhysRevLett.122.047202} {\bibfield  {journal}
  {\bibinfo  {journal} {Physical Review Letters}\ }\textbf {\bibinfo {volume}
  {122}},\ \bibinfo {pages} {047202} (\bibinfo {year} {2019})}\BibitemShut
  {NoStop}%
\bibitem [{\citenamefont {Choi}\ \emph
  {et~al.}(2019{\natexlab{a}})\citenamefont {Choi}, \citenamefont {Lee},
  \citenamefont {Lee}, \citenamefont {Yoon}, \citenamefont {Lee}, \citenamefont
  {Park}, \citenamefont {Ali}, \citenamefont {Singh}, \citenamefont {Orain},
  \citenamefont {Kim}, \citenamefont {Rhyee}, \citenamefont {Chen},
  \citenamefont {Chou},\ and\ \citenamefont {Choi}}]{choi_exotic_2019}%
  \BibitemOpen
  \bibfield  {author} {\bibinfo {author} {\bibfnamefont {Y.}~\bibnamefont
  {Choi}}, \bibinfo {author} {\bibfnamefont {C.}~\bibnamefont {Lee}}, \bibinfo
  {author} {\bibfnamefont {S.}~\bibnamefont {Lee}}, \bibinfo {author}
  {\bibfnamefont {S.}~\bibnamefont {Yoon}}, \bibinfo {author} {\bibfnamefont
  {W.-J.}\ \bibnamefont {Lee}}, \bibinfo {author} {\bibfnamefont
  {J.}~\bibnamefont {Park}}, \bibinfo {author} {\bibfnamefont {A.}~\bibnamefont
  {Ali}}, \bibinfo {author} {\bibfnamefont {Y.}~\bibnamefont {Singh}}, \bibinfo
  {author} {\bibfnamefont {J.-C.}\ \bibnamefont {Orain}}, \bibinfo {author}
  {\bibfnamefont {G.}~\bibnamefont {Kim}}, \bibinfo {author} {\bibfnamefont
  {J.-S.}\ \bibnamefont {Rhyee}}, \bibinfo {author} {\bibfnamefont {W.-T.}\
  \bibnamefont {Chen}}, \bibinfo {author} {\bibfnamefont {F.}~\bibnamefont
  {Chou}}, \ and\ \bibinfo {author} {\bibfnamefont {K.-Y.}\ \bibnamefont
  {Choi}},\ }\href {\doibase 10.1103/PhysRevLett.122.167202} {\bibfield
  {journal} {\bibinfo  {journal} {Physical Review Letters}\ }\textbf {\bibinfo
  {volume} {122}},\ \bibinfo {pages} {167202} (\bibinfo {year}
  {2019}{\natexlab{a}})}\BibitemShut {NoStop}%
\bibitem [{\citenamefont {Mehlawat}\ \emph {et~al.}(2017)\citenamefont
  {Mehlawat}, \citenamefont {Thamizhavel},\ and\ \citenamefont
  {Singh}}]{mehlawat_heat_2017}%
  \BibitemOpen
  \bibfield  {author} {\bibinfo {author} {\bibfnamefont {K.}~\bibnamefont
  {Mehlawat}}, \bibinfo {author} {\bibfnamefont {A.}~\bibnamefont
  {Thamizhavel}}, \ and\ \bibinfo {author} {\bibfnamefont {Y.}~\bibnamefont
  {Singh}},\ }\href {\doibase 10.1103/PhysRevB.95.144406} {\bibfield  {journal}
  {\bibinfo  {journal} {Physical Review B}\ }\textbf {\bibinfo {volume} {95}},\
  \bibinfo {pages} {144406} (\bibinfo {year} {2017})}\BibitemShut {NoStop}%
\bibitem [{\citenamefont {Widmann}\ \emph {et~al.}(2019)\citenamefont
  {Widmann}, \citenamefont {Tsurkan}, \citenamefont {Prishchenko},
  \citenamefont {Mazurenko}, \citenamefont {Tsirlin},\ and\ \citenamefont
  {Loidl}}]{widmann_thermodynamic_2019}%
  \BibitemOpen
  \bibfield  {author} {\bibinfo {author} {\bibfnamefont {S.}~\bibnamefont
  {Widmann}}, \bibinfo {author} {\bibfnamefont {V.}~\bibnamefont {Tsurkan}},
  \bibinfo {author} {\bibfnamefont {D.~A.}\ \bibnamefont {Prishchenko}},
  \bibinfo {author} {\bibfnamefont {V.~G.}\ \bibnamefont {Mazurenko}}, \bibinfo
  {author} {\bibfnamefont {A.~A.}\ \bibnamefont {Tsirlin}}, \ and\ \bibinfo
  {author} {\bibfnamefont {A.}~\bibnamefont {Loidl}},\ }\href {\doibase
  10.1103/PhysRevB.99.094415} {\bibfield  {journal} {\bibinfo  {journal}
  {Physical Review B}\ }\textbf {\bibinfo {volume} {99}},\ \bibinfo {pages}
  {094415} (\bibinfo {year} {2019})}\BibitemShut {NoStop}%
\bibitem [{\citenamefont {Raju}\ \emph {et~al.}(1992)\citenamefont {Raju},
  \citenamefont {Gmelin},\ and\ \citenamefont
  {Kremer}}]{raju_magnetic-susceptibility_1992}%
  \BibitemOpen
  \bibfield  {author} {\bibinfo {author} {\bibfnamefont {N.~P.}\ \bibnamefont
  {Raju}}, \bibinfo {author} {\bibfnamefont {E.}~\bibnamefont {Gmelin}}, \ and\
  \bibinfo {author} {\bibfnamefont {R.~K.}\ \bibnamefont {Kremer}},\ }\href
  {\doibase 10.1103/PhysRevB.46.5405} {\bibfield  {journal} {\bibinfo
  {journal} {Physical Review B}\ }\textbf {\bibinfo {volume} {46}},\ \bibinfo
  {pages} {5405} (\bibinfo {year} {1992})}\BibitemShut {NoStop}%
\bibitem [{\citenamefont {Mydosh}(2015)}]{mydosh_spin_2015}%
  \BibitemOpen
  \bibfield  {author} {\bibinfo {author} {\bibfnamefont {J.~A.}\ \bibnamefont
  {Mydosh}},\ }\href {\doibase 10.1088/0034-4885/78/5/052501} {\bibfield
  {journal} {\bibinfo  {journal} {Reports on Progress in Physics}\ }\textbf
  {\bibinfo {volume} {78}},\ \bibinfo {pages} {052501} (\bibinfo {year}
  {2015})}\BibitemShut {NoStop}%
\bibitem [{\citenamefont {Bette}\ \emph {et~al.}(2017)\citenamefont {Bette},
  \citenamefont {Takayama}, \citenamefont {Kitagawa}, \citenamefont {Takano},
  \citenamefont {Takagi},\ and\ \citenamefont
  {Dinnebier}}]{bette_solution_2017}%
  \BibitemOpen
  \bibfield  {author} {\bibinfo {author} {\bibfnamefont {S.}~\bibnamefont
  {Bette}}, \bibinfo {author} {\bibfnamefont {T.}~\bibnamefont {Takayama}},
  \bibinfo {author} {\bibfnamefont {K.}~\bibnamefont {Kitagawa}}, \bibinfo
  {author} {\bibfnamefont {R.}~\bibnamefont {Takano}}, \bibinfo {author}
  {\bibfnamefont {H.}~\bibnamefont {Takagi}}, \ and\ \bibinfo {author}
  {\bibfnamefont {R.~E.}\ \bibnamefont {Dinnebier}},\ }\href {\doibase
  10.1039/C7DT02978K} {\bibfield  {journal} {\bibinfo  {journal} {Dalton
  Transactions}\ }\textbf {\bibinfo {volume} {46}},\ \bibinfo {pages} {15216}
  (\bibinfo {year} {2017})}\BibitemShut {NoStop}%
\bibitem [{\citenamefont {Kenney}\ \emph {et~al.}(2019)\citenamefont {Kenney},
  \citenamefont {Segre}, \citenamefont {Lafargue-Dit-Hauret}, \citenamefont
  {Lebedev}, \citenamefont {Abramchuk}, \citenamefont {Berlie}, \citenamefont
  {Cottrell}, \citenamefont {Simutis}, \citenamefont {Bahrami}, \citenamefont
  {Mordvinova}, \citenamefont {Fabbris}, \citenamefont {McChesney},
  \citenamefont {Haskel}, \citenamefont {Rocquefelte}, \citenamefont {Graf},\
  and\ \citenamefont {Tafti}}]{kenney_coexistence_2019}%
  \BibitemOpen
  \bibfield  {author} {\bibinfo {author} {\bibfnamefont {E.~M.}\ \bibnamefont
  {Kenney}}, \bibinfo {author} {\bibfnamefont {C.~U.}\ \bibnamefont {Segre}},
  \bibinfo {author} {\bibfnamefont {W.}~\bibnamefont {Lafargue-Dit-Hauret}},
  \bibinfo {author} {\bibfnamefont {O.~I.}\ \bibnamefont {Lebedev}}, \bibinfo
  {author} {\bibfnamefont {M.}~\bibnamefont {Abramchuk}}, \bibinfo {author}
  {\bibfnamefont {A.}~\bibnamefont {Berlie}}, \bibinfo {author} {\bibfnamefont
  {S.~P.}\ \bibnamefont {Cottrell}}, \bibinfo {author} {\bibfnamefont
  {G.}~\bibnamefont {Simutis}}, \bibinfo {author} {\bibfnamefont
  {F.}~\bibnamefont {Bahrami}}, \bibinfo {author} {\bibfnamefont {N.~E.}\
  \bibnamefont {Mordvinova}}, \bibinfo {author} {\bibfnamefont
  {G.}~\bibnamefont {Fabbris}}, \bibinfo {author} {\bibfnamefont {J.~L.}\
  \bibnamefont {McChesney}}, \bibinfo {author} {\bibfnamefont {D.}~\bibnamefont
  {Haskel}}, \bibinfo {author} {\bibfnamefont {X.}~\bibnamefont {Rocquefelte}},
  \bibinfo {author} {\bibfnamefont {M.~J.}\ \bibnamefont {Graf}}, \ and\
  \bibinfo {author} {\bibfnamefont {F.}~\bibnamefont {Tafti}},\ }\href
  {\doibase 10.1103/PhysRevB.100.094418} {\bibfield  {journal} {\bibinfo
  {journal} {Physical Review B}\ }\textbf {\bibinfo {volume} {100}},\ \bibinfo
  {pages} {094418} (\bibinfo {year} {2019})}\BibitemShut {NoStop}%
\bibitem [{\citenamefont {Ramirez}(1994)}]{ramirez_strongly_1994}%
  \BibitemOpen
  \bibfield  {author} {\bibinfo {author} {\bibfnamefont {A.~P.}\ \bibnamefont
  {Ramirez}},\ }\href {\doibase 10.1146/annurev.ms.24.080194.002321} {\bibfield
   {journal} {\bibinfo  {journal} {Annual Review of Materials Science}\
  }\textbf {\bibinfo {volume} {24}},\ \bibinfo {pages} {453} (\bibinfo {year}
  {1994})}\BibitemShut {NoStop}%
\bibitem [{\citenamefont {Nasu}\ \emph {et~al.}(2015)\citenamefont {Nasu},
  \citenamefont {Udagawa},\ and\ \citenamefont {Motome}}]{nasu_thermal_2015}%
  \BibitemOpen
  \bibfield  {author} {\bibinfo {author} {\bibfnamefont {J.}~\bibnamefont
  {Nasu}}, \bibinfo {author} {\bibfnamefont {M.}~\bibnamefont {Udagawa}}, \
  and\ \bibinfo {author} {\bibfnamefont {Y.}~\bibnamefont {Motome}},\ }\href
  {\doibase 10.1103/PhysRevB.92.115122} {\bibfield  {journal} {\bibinfo
  {journal} {Physical Review B}\ }\textbf {\bibinfo {volume} {92}},\ \bibinfo
  {pages} {115122} (\bibinfo {year} {2015})}\BibitemShut {NoStop}%
\bibitem [{\citenamefont {Banerjee}\ \emph {et~al.}(2018)\citenamefont
  {Banerjee}, \citenamefont {Lampen-Kelley}, \citenamefont {Knolle},
  \citenamefont {Balz}, \citenamefont {Aczel}, \citenamefont {Winn},
  \citenamefont {Liu}, \citenamefont {Pajerowski}, \citenamefont {Yan},
  \citenamefont {Bridges}, \citenamefont {Savici}, \citenamefont {Chakoumakos},
  \citenamefont {Lumsden}, \citenamefont {Tennant}, \citenamefont {Moessner},
  \citenamefont {Mandrus},\ and\ \citenamefont
  {Nagler}}]{banerjee_excitations_2018}%
  \BibitemOpen
  \bibfield  {author} {\bibinfo {author} {\bibfnamefont {A.}~\bibnamefont
  {Banerjee}}, \bibinfo {author} {\bibfnamefont {P.}~\bibnamefont
  {Lampen-Kelley}}, \bibinfo {author} {\bibfnamefont {J.}~\bibnamefont
  {Knolle}}, \bibinfo {author} {\bibfnamefont {C.}~\bibnamefont {Balz}},
  \bibinfo {author} {\bibfnamefont {A.~A.}\ \bibnamefont {Aczel}}, \bibinfo
  {author} {\bibfnamefont {B.}~\bibnamefont {Winn}}, \bibinfo {author}
  {\bibfnamefont {Y.}~\bibnamefont {Liu}}, \bibinfo {author} {\bibfnamefont
  {D.}~\bibnamefont {Pajerowski}}, \bibinfo {author} {\bibfnamefont
  {J.}~\bibnamefont {Yan}}, \bibinfo {author} {\bibfnamefont {C.~A.}\
  \bibnamefont {Bridges}}, \bibinfo {author} {\bibfnamefont {A.~T.}\
  \bibnamefont {Savici}}, \bibinfo {author} {\bibfnamefont {B.~C.}\
  \bibnamefont {Chakoumakos}}, \bibinfo {author} {\bibfnamefont {M.~D.}\
  \bibnamefont {Lumsden}}, \bibinfo {author} {\bibfnamefont {D.~A.}\
  \bibnamefont {Tennant}}, \bibinfo {author} {\bibfnamefont {R.}~\bibnamefont
  {Moessner}}, \bibinfo {author} {\bibfnamefont {D.~G.}\ \bibnamefont
  {Mandrus}}, \ and\ \bibinfo {author} {\bibfnamefont {S.~E.}\ \bibnamefont
  {Nagler}},\ }\href {\doibase 10.1038/s41535-018-0079-2} {\bibfield  {journal}
  {\bibinfo  {journal} {npj Quantum Materials}\ }\textbf {\bibinfo {volume}
  {3}},\ \bibinfo {pages} {1} (\bibinfo {year} {2018})}\BibitemShut {NoStop}%
\bibitem [{\citenamefont {Yaouanc}\ and\ \citenamefont {Dalmas~de
  R\'{e}otier}(2011)}]{Yaouanc_muon_2011}%
  \BibitemOpen
  \bibfield  {author} {\bibinfo {author} {\bibfnamefont {A.}~\bibnamefont
  {Yaouanc}}\ and\ \bibinfo {author} {\bibfnamefont {P.}~\bibnamefont
  {Dalmas~de R\'{e}otier}},\ }\href@noop {} {\emph {\bibinfo {title} {Muon
  {Spin} {Rotation}, {Relaxation}, and {Resonance}: {Applications} to
  {Condensed} {Matter}}}},\ International {Series} of {Monographs} on
  {Physics}\ (\bibinfo  {publisher} {Oxford University Press},\ \bibinfo
  {address} {Oxford, New York},\ \bibinfo {year} {2011})\BibitemShut {NoStop}%
\bibitem [{\citenamefont {Kao}\ \emph {et~al.}(2020)\citenamefont {Kao},
  \citenamefont {Knolle}, \citenamefont {Halász}, \citenamefont {Moessner},\
  and\ \citenamefont {Perkins}}]{kao_vacancy-induced_2020}%
  \BibitemOpen
  \bibfield  {author} {\bibinfo {author} {\bibfnamefont {W.-H.}\ \bibnamefont
  {Kao}}, \bibinfo {author} {\bibfnamefont {J.}~\bibnamefont {Knolle}},
  \bibinfo {author} {\bibfnamefont {G.~B.}\ \bibnamefont {Halász}}, \bibinfo
  {author} {\bibfnamefont {R.}~\bibnamefont {Moessner}}, \ and\ \bibinfo
  {author} {\bibfnamefont {N.~B.}\ \bibnamefont {Perkins}},\ }\href
  {http://arxiv.org/abs/2007.11637} {\bibfield  {journal} {\bibinfo  {journal}
  {arXiv:2007.11637 [cond-mat]}\ } (\bibinfo {year} {2020})}\BibitemShut
  {NoStop}%
\bibitem [{\citenamefont {Abramchuk}\ \emph {et~al.}(2018)\citenamefont
  {Abramchuk}, \citenamefont {Lebedev}, \citenamefont {Hellman}, \citenamefont
  {Bahrami}, \citenamefont {Mordvinova}, \citenamefont {Krizan}, \citenamefont
  {Metz}, \citenamefont {Broido},\ and\ \citenamefont
  {Tafti}}]{abramchuk_crystal_2018}%
  \BibitemOpen
  \bibfield  {author} {\bibinfo {author} {\bibfnamefont {M.}~\bibnamefont
  {Abramchuk}}, \bibinfo {author} {\bibfnamefont {O.~I.}\ \bibnamefont
  {Lebedev}}, \bibinfo {author} {\bibfnamefont {O.}~\bibnamefont {Hellman}},
  \bibinfo {author} {\bibfnamefont {F.}~\bibnamefont {Bahrami}}, \bibinfo
  {author} {\bibfnamefont {N.~E.}\ \bibnamefont {Mordvinova}}, \bibinfo
  {author} {\bibfnamefont {J.~W.}\ \bibnamefont {Krizan}}, \bibinfo {author}
  {\bibfnamefont {K.~R.}\ \bibnamefont {Metz}}, \bibinfo {author}
  {\bibfnamefont {D.}~\bibnamefont {Broido}}, \ and\ \bibinfo {author}
  {\bibfnamefont {F.}~\bibnamefont {Tafti}},\ }\href {\doibase
  10.1021/acs.inorgchem.8b01866} {\bibfield  {journal} {\bibinfo  {journal}
  {Inorganic Chemistry}\ }\textbf {\bibinfo {volume} {57}},\ \bibinfo {pages}
  {12709} (\bibinfo {year} {2018})}\BibitemShut {NoStop}%
\bibitem [{\citenamefont {Rousochatzakis}\ \emph {et~al.}(2019)\citenamefont
  {Rousochatzakis}, \citenamefont {Kourtis}, \citenamefont {Knolle},
  \citenamefont {Moessner},\ and\ \citenamefont
  {Perkins}}]{rousochatzakis_quantum_2019}%
  \BibitemOpen
  \bibfield  {author} {\bibinfo {author} {\bibfnamefont {I.}~\bibnamefont
  {Rousochatzakis}}, \bibinfo {author} {\bibfnamefont {S.}~\bibnamefont
  {Kourtis}}, \bibinfo {author} {\bibfnamefont {J.}~\bibnamefont {Knolle}},
  \bibinfo {author} {\bibfnamefont {R.}~\bibnamefont {Moessner}}, \ and\
  \bibinfo {author} {\bibfnamefont {N.~B.}\ \bibnamefont {Perkins}},\ }\href
  {\doibase 10.1103/PhysRevB.100.045117} {\bibfield  {journal} {\bibinfo
  {journal} {Physical Review B}\ }\textbf {\bibinfo {volume} {100}},\ \bibinfo
  {pages} {045117} (\bibinfo {year} {2019})}\BibitemShut {NoStop}%
\bibitem [{\citenamefont {Choi}\ \emph
  {et~al.}(2019{\natexlab{b}})\citenamefont {Choi}, \citenamefont {Manni},
  \citenamefont {Singleton}, \citenamefont {Topping}, \citenamefont
  {Lancaster}, \citenamefont {Blundell}, \citenamefont {Adroja}, \citenamefont
  {Zapf}, \citenamefont {Gegenwart},\ and\ \citenamefont
  {Coldea}}]{choi_spin_2019}%
  \BibitemOpen
  \bibfield  {author} {\bibinfo {author} {\bibfnamefont {S.}~\bibnamefont
  {Choi}}, \bibinfo {author} {\bibfnamefont {S.}~\bibnamefont {Manni}},
  \bibinfo {author} {\bibfnamefont {J.}~\bibnamefont {Singleton}}, \bibinfo
  {author} {\bibfnamefont {C.~V.}\ \bibnamefont {Topping}}, \bibinfo {author}
  {\bibfnamefont {T.}~\bibnamefont {Lancaster}}, \bibinfo {author}
  {\bibfnamefont {S.~J.}\ \bibnamefont {Blundell}}, \bibinfo {author}
  {\bibfnamefont {D.~T.}\ \bibnamefont {Adroja}}, \bibinfo {author}
  {\bibfnamefont {V.}~\bibnamefont {Zapf}}, \bibinfo {author} {\bibfnamefont
  {P.}~\bibnamefont {Gegenwart}}, \ and\ \bibinfo {author} {\bibfnamefont
  {R.}~\bibnamefont {Coldea}},\ }\href {\doibase 10.1103/PhysRevB.99.054426}
  {\bibfield  {journal} {\bibinfo  {journal} {Physical Review B}\ }\textbf
  {\bibinfo {volume} {99}},\ \bibinfo {pages} {054426} (\bibinfo {year}
  {2019}{\natexlab{b}})}\BibitemShut {NoStop}%
\bibitem [{\citenamefont {Pratt}(2007)}]{pratt_field_2007}%
  \BibitemOpen
  \bibfield  {author} {\bibinfo {author} {\bibfnamefont {F.~L.}\ \bibnamefont
  {Pratt}},\ }\href {\doibase 10.1088/0953-8984/19/45/456207} {\bibfield
  {journal} {\bibinfo  {journal} {Journal of Physics: Condensed Matter}\
  }\textbf {\bibinfo {volume} {19}},\ \bibinfo {pages} {456207} (\bibinfo
  {year} {2007})}\BibitemShut {NoStop}%
\end{thebibliography}%
%%%%%%%%%%%%%%%%%%%%%%%%%%%%%%%%%%%%%%%%%%%%%%%

\end{document}